\documentclass{mn2e}

\usepackage{amsfonts}
\usepackage{amsmath}
\usepackage{graphicx}

\title[Stellar Mass Fundamental Plane]
      {The luminosity and stellar mass Fundamental Plane of early-type galaxies}

\author[J. B. Hyde \& M. Bernardi]
{Joseph B. Hyde \& Mariangela Bernardi\thanks{E-mail: jhyde,bernardm@physics.upenn.edu}\\
      Department of Physics \& Astronomy, University of Pennsylvania, 
      209 S. 33rd St., Philadelphia, PA 19104, USA}

\begin{document}
\pagerange{\pageref{firstpage}--\pageref{lastpage}}

\maketitle

\label{firstpage}

\begin{abstract}
From a sample of $\sim 50,000$ early-type galaxies from the Sloan Digital
Sky Survey (SDSS), we 
measured the traditional Fundamental Plane in the $g$, $r$, $i$ and $z$ 
bands.  We then replaced luminosity with stellar mass, and measured 
the ``stellar mass'' Fundamental Plane. 
The Fundamental Plane, $R\propto \sigma^a/I^B$, steepens 
slightly as one moves from shorter to longer wavelengths:  
the orthogonal fit has slope $a=1.40$ in the $g$ band and $1.47$ in $z$, 
with a statistical random error of $\sim 0.02$.
However, systematic effects can produce larger uncertainties, of 
order $\sim 0.05$.
The Fundamental Plane is thinner at longer wavelengths: 
it has an intrinsic scatter of 0.062~dex in $g$ and 0.054~dex in $z$.
We have clear evidence that the scatter is larger at small galaxy 
sizes/masses; at large masses measurement errors account for essentially 
all of the observed scatter (about 0.04~dex), suggesting that the Plane is 
rather thin for the very massive galaxies.  
The Fundamental Plane steepens further when luminosity is replaced with 
stellar mass, to $1.54$ or $1.63$ when stellar masses are estimated from 
broad-band colors or from spectra, respectively.  
The intrinsic scatter also reduces further, to 0.048~dex on average.
Since color and stellar mass-to-light ratio are closely related, 
this explains why color can be thought of as the fourth Fundamental 
Plane parameter.  However, the slope of the stellar mass Fundamental 
Plane remains shallower than the value of 2 associated with the 
virial theorem.  This is because the ratio of dynamical to stellar mass 
increases at large masses: 
 $M_{dyn}/M_*\propto M_{dyn}^{0.17\pm0.01}$.  This scaling is the edge-on 
projection of the stellar mass $\kappa$-space.  The face-on view 
suggests that there is an upper limit to the stellar density for a 
given dynamical mass, and this decreases at large masses:
  $M_*/R_e^3 \propto M_{dyn}^{-4/3}$.
All these trends can be used to constrain early-type galaxy formation 
models.  

We also study how the estimated coefficients $a$ and $B$ of the
Plane are affected by other selection effects, whether in apparent 
or absolute quantities.  For example, if low luminosity objects 
are missing from the sample, and one does not account for this, 
then $a$ and $B$ are both biased low from their true values.  
If objects with small velocity dispersions are missing, then $a$ 
is biased high, although this matters more for the orthogonal 
than the direct-fitted quantities.  These biases are seen in 
Fundamental Planes which have no intrinsic curvature, so the 
observation that $a$ and $B$ scale with $L$ and $\sigma$ is not, 
by itself, evidence that the Plane is warped.  On the other hand, 
we show that the Plane appears to curve sharply downwards at the 
small-size/mass end, and more gradually downwards as one moves 
towards larger sizes/masses.  Whereas the drop at small sizes is 
real, most of the latter effect is due to correlated errors.  
\end{abstract}


\section{\protect\bigskip Introduction}
The Fundamental Plane has been a useful diagnostic of galaxy 
distances and galaxy evolution (e.g. Dressler et al. 1987; 
Djorgovski \& Davis 1987; J{\o}rgensen et al. 1996; 
Pahre et al. 1998; Bernardi et al. 2003; J{\o}rgensen et al. 2007).  
If galaxies are virialized, then one expects observable signatures 
of the balance between potential and kinetic energies.  For example, 
\begin{equation}
 \sigma^2 \propto \frac{GM_{dyn}}{R} 
          \propto \frac{M_{dyn}}{L}\,\frac{L}{R^2}\, R
          \propto \frac{M_{dyn}}{L}\,I\,R,
 \label{vir}
\end{equation}
where $\sigma$ is a velocity dispersion, $I$ is a surface 
brightness, $R$ is a scale, and $M_{dyn}/L$ is the mass to light ratio.  
This suggests that the observed line-of-sight velocity dispersion 
$\sigma_e^2$ and surface brightness $I_e\equiv L/(2 \pi R_e^2)$ at or 
within some fiducial radius $R_e$ should be correlated with 
one another.  (In what follows, we will use the subscript $e$ 
to denote quantities estimated from deVaucouleur's (1948) fits to 
the surface brightness profiles.)

Velocity dispersion and surface brightness are distance-independent 
observables, whereas the size is not, so it is common to write 
\begin{equation}
 R_e \propto \sigma_e^a \, I_e^{-B},
\end{equation}
and to find that pair $(a,B)$ for which the scatter either in the 
$R_e$ direction, or in the direction orthogonal to the Plane, is 
minimized.  If the resulting values of $(a,B)$ differ from the 
virial scalings $(2,1)$, then the Fundamental Plane is said to be 
`tilted'.  The tilt of the Fundamental Plane is interpreted as 
evidence that the mass-to-light ratio $(M_{dyn}/L)$ depends on 
some combination of the observables $(R_e,I_e,\sigma_e)$ -- the 
prejudice that galaxies are virialized sets what this combination 
must be:
  $(M_{dyn}/L)\propto \sigma_e^{2-a}\,I_e^{B-1}$.  
In low redshift ($z\sim 0.1$) samples, $a\sim 1.5$ and $B\sim 0.8$, 
so the implied $M_{dyn}/L$ is expected to increase slightly with $M_{dyn}$ 
or $L$.  J{\o}rgensen et al. (2007) find that $a=0.6$, $B=0.7$ 
and $M_{dyn}/L\propto M^{0.54}$ at $z\sim 0.9$.  

Contributions to the tilt of the Fundamental Plane can be further 
examined by expressing the mass-to-light ratio in terms of the dynamical 
mass, total mass, stellar mass, and broadband luminosity:
\begin{equation}
\label{e_ml}
 \left(\frac{M_{dyn}}{L}\right)\ = \left(\frac{M_{dyn}}{M_{tot}}\right)\ 
  \left(\frac{M_{tot}}{M_*}\right)\ \left(\frac{M_*}{L_{}}\right)\
\end{equation}
where $M_{tot}$ is the sum of dark matter and baryonic mass, 
$M_*$ is the stellar mass, and $L$ is the observed broadband luminosity.
If any of these terms varies as a function of $M_{dyn}$ or $L$, the Plane 
will be tilted. 

The assumption that $M_{dyn}/M_{tot}$ is constant is the assumption of 
homology. The validity of this assumption is supported by detailed 
kinematic modeling of galaxies observed with SAURON integral-field 
spectroscopy.  The combination of 2-D photometric and spectroscopic data 
allowed Cappellari et al. (2007) to account for differences in 
e.g. $\rho(r)$, $\sigma(r)$, and $V(r)$.  They found that dynamical 
mass is a robust tracer of total mass.  More recent work by 
Bolton et al. (2008) suggests that the deVaucouleur-based quantity 
$R_e\sigma_e^2$ is linearly proportional to the mass estimated from the 
strong gravitational lensing effect, further reinforcing the homology 
assumption that $M_{dyn}\propto M_{tot}$.

For a given IMF, the stellar mass to light ratio, $(M_*/L)$, depends 
on the age and metallicity of the stellar population as well as on 
wavelength (e. g., Tinsley 1978; Worthey 1994).  
There is now growing evidence for correlations between mass and 
metallicity (e.g., Trager et al. 2000; Nelan et al. 2005; 
Thomas et al. 2005), and between age and velocity dispersion 
(e.g., Bernardi et al. 2005).  
Hence, stellar population models predict that $(M_*/L)$ will depend 
on total mass and luminosity, plausibly producing a tilt that will 
depend on waveband.   While there is evidence that the tilt does 
indeed depend on waveband, the dependence is weak 
(Pahre et al. 1998; Bernardi et al. 2003; La Barbera et al. 2008), 
and so this effect alone cannot explain all of the tilt.  

There is little discussion in the literature of the possibility 
that the IMF changes along the sequence in such a way as to make 
the dependence on $M_*/L$ weak.  If there is no such effect, 
then stellar population effects and non-homology are weak, so the 
dominating contribution to the tilt of the Fundamental Plane is due 
to the systematic variation of $M_{tot}/M_*$ with mass.  
In a hierarchical scenario of galaxy formation, the relative distribution 
of stars and dark matter in a galaxy depend on the roles of dissipational 
and dissipationless merging (e.g. Bender et al. 1992). Hydrodynamical 
simulations of galaxy mergers (Robertson et al. 2006; Hopkins et al. 2008) 
suggest that the fractional gas content of merging galaxies determines the 
fundamental Plane of their remnants.  Further dissipationless mergers 
preserve the Fundamental Plane, but not its projections 
(e.g., Boylan-Kolchin et al. 2005).  Observations of brightest cluster galaxies by
Bernardi et al. (2007) support these conclusions.

The dissipational content in mergers sets the effective mass-to-light 
ratio of the merger remnant (Robertson et al. 2006; Hopkins et al. 2008). 
If the dissipational content varies with mass (in spirals the gas 
fraction decreases with mass, e.g. Bell \& deJong 2000), this would 
cause a systematic dependence of the mass-to-light ratio with mass.  
If this ratio varies as a power law with mass ($M/L\propto M^\alpha$), 
then the Fundamental Plane would be tilted relative to the virial scaling. 
If the mass-to-light ratio varies in a more complicated manner, the 
Fundamental Plane could be warped, or have a more complicated shape.

Therefore, in the present work, we separate out the contribution from 
these effects by re-writing equation~(\ref{vir}) as:
\begin{equation}
 \sigma^2 \propto \left(\frac{M_{dyn}}{M_*}\right)\,
                  \left(\frac{M_*}{L}\right)\,
                  \left(\frac{L}{R^2}\right)\, R.
\end{equation}
We do not consider variation of $(M_{dyn}/M_{tot})$, since two rather 
different methods (Cappellari et al. 2007; Bolton et al. 2008) suggest 
that this ratio is a constant across the early-type population.  
(This is also why we do not include effects associated with fitting 
Sersic's 1968 generalization of the deVaucouleur profile to the 
images, even though there is a well-developed literature on this 
subject.)  
If the stellar population models which one uses to estimate $M_*/L$
are accurate, then multiplying the surface brightness by $M_*/L$ 
should eliminate most of the dependence on waveband.  (There is a 
small remaining dependence which arises from the fact that the 
half-light radii are slightly but systematically smaller in redder 
bands, e.g. Hyde \& Bernardi 2009).  
The remaining term $(M_{dyn}/M_*)$ is the ratio of the dynamical to 
stellar mass, which we will use as our proxy for structural 
differences.  This is interesting because $M_{dyn}/M_*$ is not 
expected to be very much greater than unity -- recent work 
suggests that $M_{dyn}/M_* \approx 1/0.7$ (Capellari et al. 2007) -- 
how it scales with mass is a quantity of great interest to galaxy 
formation models (e.g. Bower et al. 2006; DeLucia et al. 2006; 
Hopkins et al. 2008), and previous work in the SDSS suggests 
that $M_{dyn}/M_*$ is not constant across the early-type population 
(Padmanabhan et al. 2004; Gallazzi et al. 2006)

Section~\ref{FPtrad} discusses how we select our sample and 
shows the result of fitting for the coefficients $(a,B)$ of the 
traditional Fundamental Plane in the SDSS $g$, $r$, $i$, and $z$ bands.  
(Appendix~A shows that our sample probably contains some 
non-early-type galaxies but they are too few to affect our conclusions.)
We also study the possibility of detecting if the Plane is warped, 
or merely thick.  This is prompted in part by recent work showing 
that the parameters of the fitted Fundamental Plane appear to depend 
on the luminosity and velocity dispersion range of the sample 
(D'Onofrio et al. 2008; Nigoche-Netro et al. 2009).  We show that 
such dependences exist in thick Planes that are not warped -- they 
are a consequence of not accounting for selection effects.  We then 
discuss more reliable estimates of curvature along the Plane.  

Section~\ref{FPmass} describes our estimates of $M_*/L$ and the 
result of fitting for the coefficients $(\alpha,\beta)$ in 
\begin{equation}
 R_e \propto \sigma_e^\alpha \, \Sigma_e^{-\beta}
\end{equation}
where $\Sigma \equiv (M_*/L)\,I$.  The difference between 
$(\alpha,\beta)$ and $(a,B)$ is due to stellar population 
effects; the difference between $(\alpha,\beta)$ and the 
virial scalings $(2,1)$ encodes information about $M_{dyn}/M_*$.  
If $(\alpha,\beta)<(2,1)$ then $(M_{dyn}/M_*)$ increases 
with $M_{dyn}$ or $M_*$ for the same reason that $(a,B)<(2,1)$ 
implies $(M_{dyn}/L)$ increases with $M_{dyn}$ or $L$.  

A complementary analysis of the Fundamental Plane variables was 
introduced by Bender, Burstein \& Faber (1992).  This construction, 
which they termed $\kappa$-space, was criticized as being an 
``obfuscation, not a simplification'', by Pahre et al. (1998), 
primarily on the grounds that $R_e$ and $I_e$ depend on waveband.  
However, by replacing luminosity with stellar mass, what we will 
call $\kappa_*$-space, most of the wavelength dependence is removed; 
only the weak dependence of $R_e$ on waveband remains.  
So it is interesting to examine how early-types are distributed in 
$\kappa_*$-space.  Section~\ref{kspace} presents the first analysis 
of $\kappa_*$-space.  The scaling of $M_*/M_{dyn}$ with $M_{dyn}$ 
referred above is the edge-on view of the $\kappa_*$-Plane.  The 
face-on view provides a relation between the maximum stellar density 
and dynamical mass of early-type galaxies.  

A final section summarizes our findings and discusses some 
implications.

\section{FP:  The Traditional Fundamental Plane}\label{FPtrad}
We use the sample of about 50,000 early-type galaxies assembled 
by Hyde \& Bernardi (2008).  The sample is based on the SDSS-DR4, 
but with photometric and spectroscopic parameters updated from the 
SDSS-DR6 database.  
Briefly, objects have deVaucouleur magnitudes $14.5<m_r<17.5$, 
${\tt fracDev = 1}$ in both the $g$- and $r$-bands,  
and axis ratios $b/a\ge 0.6$.  
Appendix~A shows that these cuts almost certainly do not yield 
a pure early-type galaxy sample (e.g., the face-on analogues of 
the objects with {\tt fracDev=1} and $b/a\le 0.6$ are still in 
our sample).  However, the non-early types in our sample are too 
few to affect our conclusions.

Photometric parameters for the best-fit deVaucouleur surface 
brightness profiles are ``corrected'' for known problems which 
arise from the SDSS sky-subtraction algorithm.  
In addition, there are some systematic differences between the 
velocity dispersions output by SDSS-DR6 and the {\tt IDLspec2d} 
reduction (Hyde \& Bernardi 2009).  Our velocity dispersion 
estimates are simply the average of the two reductions (Section~\ref{errors}
discusses the dependence of the Fundamental Plane parameters on systematics).
We also compute the velocity dispersion for objects for which the SDSS 
pipeline does not estimate $\sigma$ (due to low $S/N$ or the presence 
of weak emission lines, i.e. the {\tt status} flag not-equal to 4).
We select galaxies with velocity dispersions $60 < \sigma < 400$~km~s$^{-1}$.  
The velocity dispersions are then corrected to $R_e/8$.  
Stellar mass estimates, from Gallazzi et al. (2005), are available for 
all these objects.  Since these are actually stellar mass to light 
ratios multiplied by a luminosity, and we have corrected the luminosities 
for the sky-subtraction problems whereas Gallazzi et al. did not, 
we apply a correction to their stellar masses to account for this effect.
Where necessary, distances were computed from redshifts assuming 
a Hubble constant of $70$~km~s${-1}$~Mpc$^{-1}$ in a flat 
$\Lambda$CDM model with $\Omega_0=0.3$.

\subsection{Fitting the Plane}
We fit the Plane as follows.
We begin by writing the Fundamental Plane as 
\begin{equation}
 \log_{10}\left(\frac{R_e}{{\rm kpc}}\right) = a\,\log_{10}\left(\frac{\sigma}{{\rm km~s^{-1}}}\right)
                          + b\,\frac{\mu_e}{{\rm mags}} + c,
\end{equation}
where $c = \langle \log_{10}R\rangle - a\,\langle\log_{10}\sigma\rangle 
                               - b\,\langle\mu_e\rangle$, and $\mu_e$ is the mean surface brightness within the half light 
radius, defined explicitly as follows:
\begin{eqnarray}
\mu_e &=& -2.5\log_{10}(I_e) = -2.5\log_{10} \left(\frac{L}{2\pi R_e^2}\right) \nonumber\\ 
&=& m + 5\log_{10}(r_e)+2.5\log_{10}(2\pi) - 10\log_{10}(1+z)
\end{eqnarray}
where $m$ is the evolution, reddening, and k-corrected apparent magnitude
(refer to Hyde \& Bernardi 2009 for details on magnitudes),
and $r_e$ is the angular size in arcseconds. 
In this form, the virial scaling would follow $(a,b)=(2,0.4)$
because of the $-2.5$ term in the definition of $\mu_e$.
Unless we say so explicitly, we work in logarithmic units.  
Therefore, following Bernardi et al. (2003), we will sometimes 
abuse notation by using
 $R$ to denote $\log_{10} (R_e$/kpc), 
 $V$ to denote $\log_{10} (\sigma$/km~s$^{-1}$), and 
 $I$ to denote $\mu_e$.
Thus, for example,
 $e_R, e_V$ and $e_I$ denote the measurement errors on
 $\log R_e$, $\log \sigma$, and $\mu_e$.

The shape of the Fundamental Plane is determined by estimating 
$a$ and $b$.  This is done either by minimizing residuals in the 
$R_e$ direction, or in the direction orthogonal to the fit.  
In general the `direct' and `orthogonal' fit parameters are 
different combinations of the mean values of and covariances 
between the variables $\log_{10}R$, $\log_{10}\sigma$ and $\mu_e$.

In practice, naive estimation of these means and covariances 
(e.g. simply summing over the data without including other 
weight terms) may lead to biases induced by measurement errors 
(these usually affect the covariances) or by selection effects 
(which bias the means and the covariances).  
The effects of both must be accounted-for to estimate the 
intrinsic shape parameters $a$ and $b$ (e.g. Saglia et al. 2001).  
This is especially important when the FP is determined for galaxies 
in a magnitude limited sample (Bernardi et al. 2003).  
We do this following methods described in Sheth \& Bernardi (2009). 
Analytic expressions which quantify the bias in the 
Fundamental Plane due to the magnitude limit of the survey 
may be found there.  

Briefly, the effect of the magnitude limit is removed by weighting 
each object by $V_{\rm max}^{-1}(L)$, the inverse of the volume 
over which it could have been observed, and measurement errors 
are subtracted in quadrature from the measured covariances.  
The mean values of magnitude, size, surface-brightness, and 
velocity dispersion in the $r$-band are 
\begin{eqnarray}
 \langle M_r/{\rm mags} \rangle &=& -20.99,\qquad\ \ \,
 \langle \log R_e/{\rm kpc}\rangle = 0.39, \nonumber\\ 
 \langle \mu_e\rangle &=&\ \ 19.53,\qquad
 \langle \log \sigma/{\rm km~s}^{-1}\rangle = 2.19.
\end{eqnarray}
With typical measurement errors 
\begin{eqnarray}
  {\cal E} &=& 
 \left(\begin{array}{ccc}
  \langle e_I^2\rangle & \langle e_I e_R\rangle & \langle e_I e_V\rangle \\
  \langle e_I e_R\rangle & \langle e_R^2\rangle & \langle e_R e_V\rangle \\
  \langle e_I e_V\rangle & \langle e_R e_V\rangle & \langle e_V^2\rangle \\
          \end{array}\right) \nonumber\\
 &=& 
 \left(\begin{array}{ccc}
           0.0542  & 0.0162 &  0 \\
           0.0162  & 0.0049 &  0 \\
           0       & 0      &  0.0016 \\
          \end{array}\right)
\end{eqnarray}
we find that the intrinsic covariance matrix (i.e. corrected for 
measurement errors and the magnitude limited selection effect) is 
\begin{eqnarray}
 {\cal F} &=&
 \left(\begin{array}{ccc}
           C_{II} & C_{IR} & C_{IV}\\
           C_{IR} & C_{RR} & C_{RV}\\
           C_{IV} & C_{RV} & C_{VV}\\
          \end{array}\right) \nonumber\\
 &=& 
 \left(\begin{array}{rcr}
           0.2947  & 0.0782 & -0.0096\\
           0.0782  & 0.0552 &  0.0189\\
          -0.0096  & 0.0204 &  0.0187 \\
          \end{array}\right).
\end{eqnarray}
From this covariance matrix we obtain 
\begin{eqnarray}
 a_{\rm dir} &=& 1.170 \quad {\rm and}\quad 
 b_{\rm dir} = 0.303 \\
 a_{\rm ort} &=& 1.434 \quad {\rm and}\quad
 b_{\rm ort} = 0.315 .
\end{eqnarray}
The observed rms scatter about the direct fit is 0.107, of which 
0.096 is intrinsic.  These quantities for the orthogonal fit are 
0.066 and 0.058 respectively. The uncertainty on the coefficients $a$ and
$b$ is dominated by systematic effects more than random errors, as described
here below. Typical uncertainties on the coefficients due to random errors 
are $\delta a < 0.02$ and $\delta b < 0.01$. Sytematics errors 
give $\delta a \sim 0.05$ and $\delta b \sim 0.02$.

\subsubsection{Dependence of $\delta a$ and $\delta b$ on sample size}
\label{samplesize}

\begin{figure}
 \centering
 \includegraphics[width=\hsize]{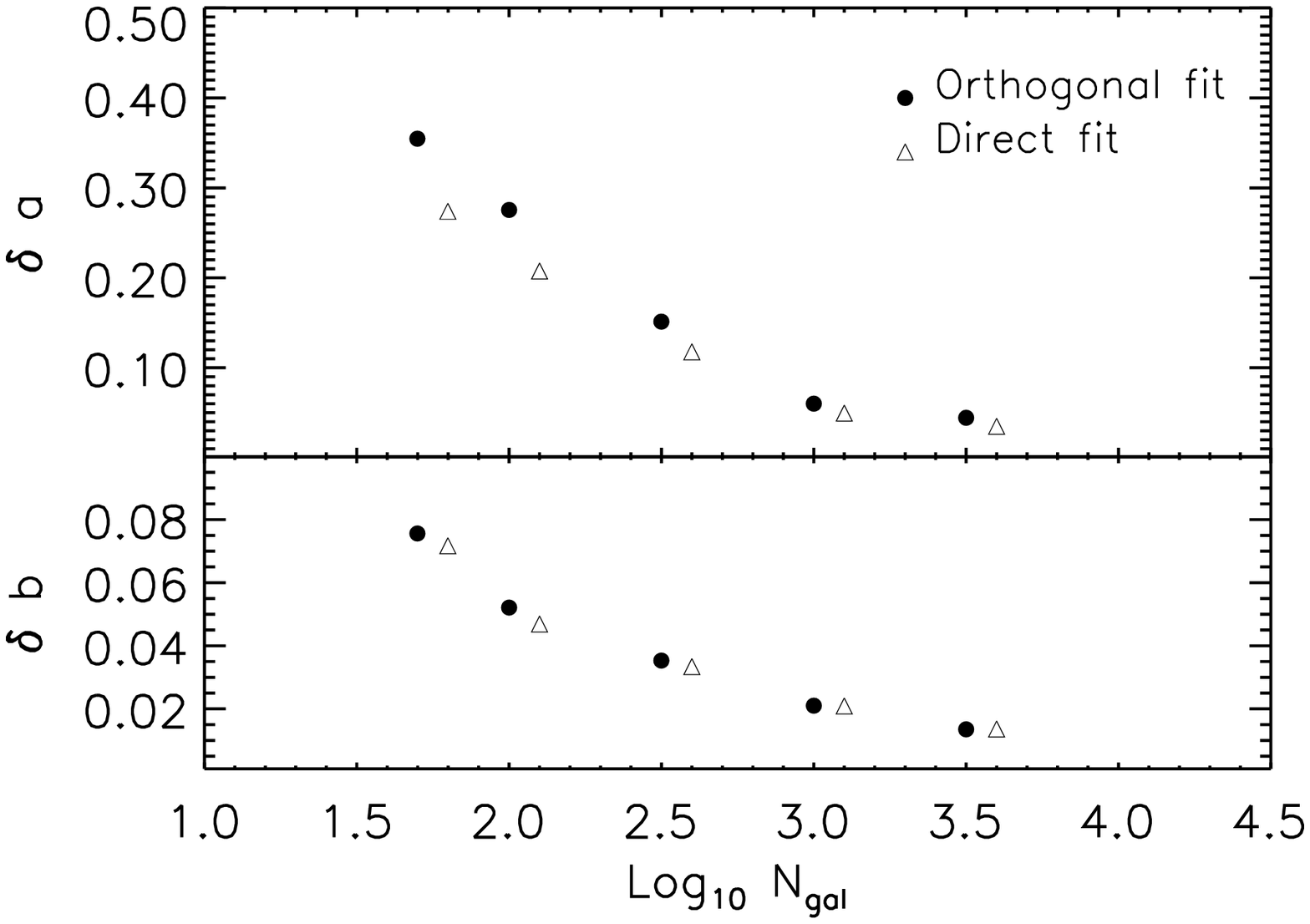}
 \includegraphics[width=0.95\hsize]{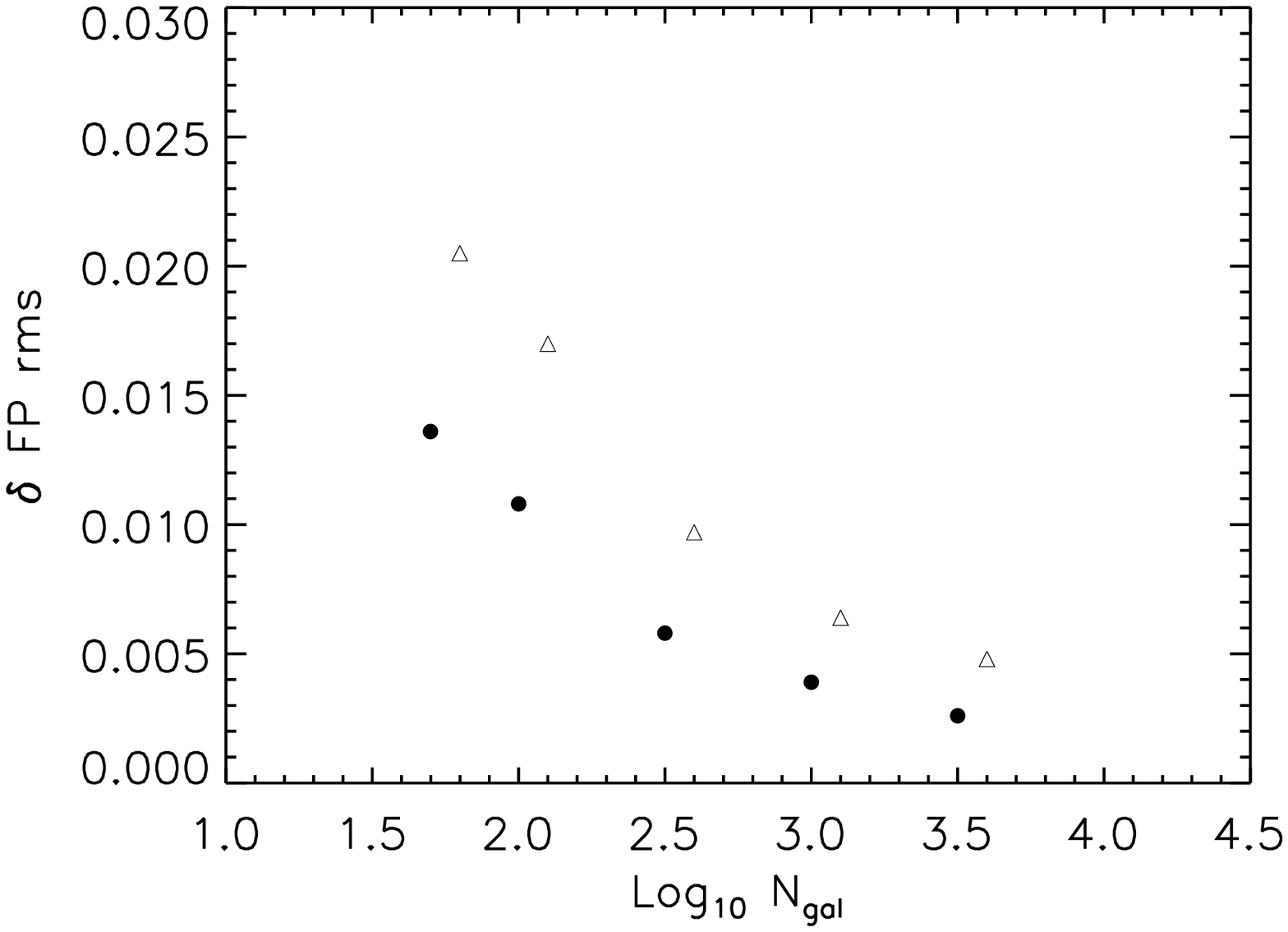}
 \caption{Precision of the FP coefficients $a$, $b$ and scatter as the 
          number of galaxies used to fit the FP changes. 
          The quantities $\delta a$, $\delta b$ and $\delta$FPrms show 
          the standard deviation values of $a$, $b$ and the scatter around the FP, 
          obtained from fitting the FP to $N=N_{\rm tot}/N_{\rm gal}$ 
          subsamples of the whole sample (i.e. each galaxy was chosen 
          only once).  Triangles and circles show results for the 
          direct and orthogonal fits, respectively. For clarity, the 
          triangles have been slightly shifted to the right. }
 \label{err_coef}
\end{figure}

\begin{figure}
 \centering
 \includegraphics[width=0.95\hsize]{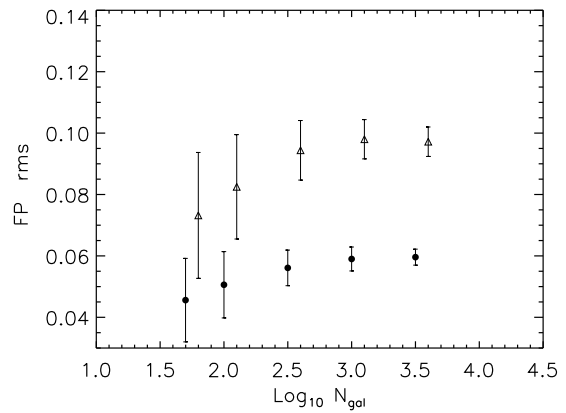}
 \caption{As for Figure~\ref{err_coef}, but now for the dependence of the 
          estimated intrinsic scatter around the FP on sample size.
          Error bars show the $\delta$FPrms values from Figure~\ref{err_coef}.}
 \label{err_rms}
\end{figure}

We analyze how the uncertainty on $a$ and $b$ depend on sample size using
bootstrap resampling. Figures~\ref{err_coef} and~\ref{err_rms} show how 
$\delta a$ and $\delta b$ depend on sample size.  They were obtained by 
dividing the total sample up into smaller subsamples, each containing 
$N_{\rm gal}$ galaxies.  For a given $N_{\rm gal}$, we do not include any 
galaxy more than once, so there are a total of $N = N_{\rm tot}/N_{\rm gal}$ 
subsamples.  We then fit the FP in each subsample, and computed the 
mean and rms values of $a$, $b$ and intrinsic scatter.  
Figure~\ref{err_coef} shows that the precision increases as the 
number of objects in the sample increases, as one might expect.  
However, Figure~\ref{err_rms} shows that sample sizes smaller than 
about 300 tend to result in underestimates of the intrinsic scatter, 
in good agreement with La Barbera et al. (2000).  
For a sample of $\sim 50,000$ galaxies the statistical random error 
$\delta a$ and $\delta b$ are quite small: $\delta a < 0.02$ and 
$\delta b < 0.01$.

\subsubsection{Dependence of $\delta a$ and $\delta b$ on systematics}
\label{errors}

\begin{figure*}
 \centering
 \includegraphics[width=0.475\hsize]{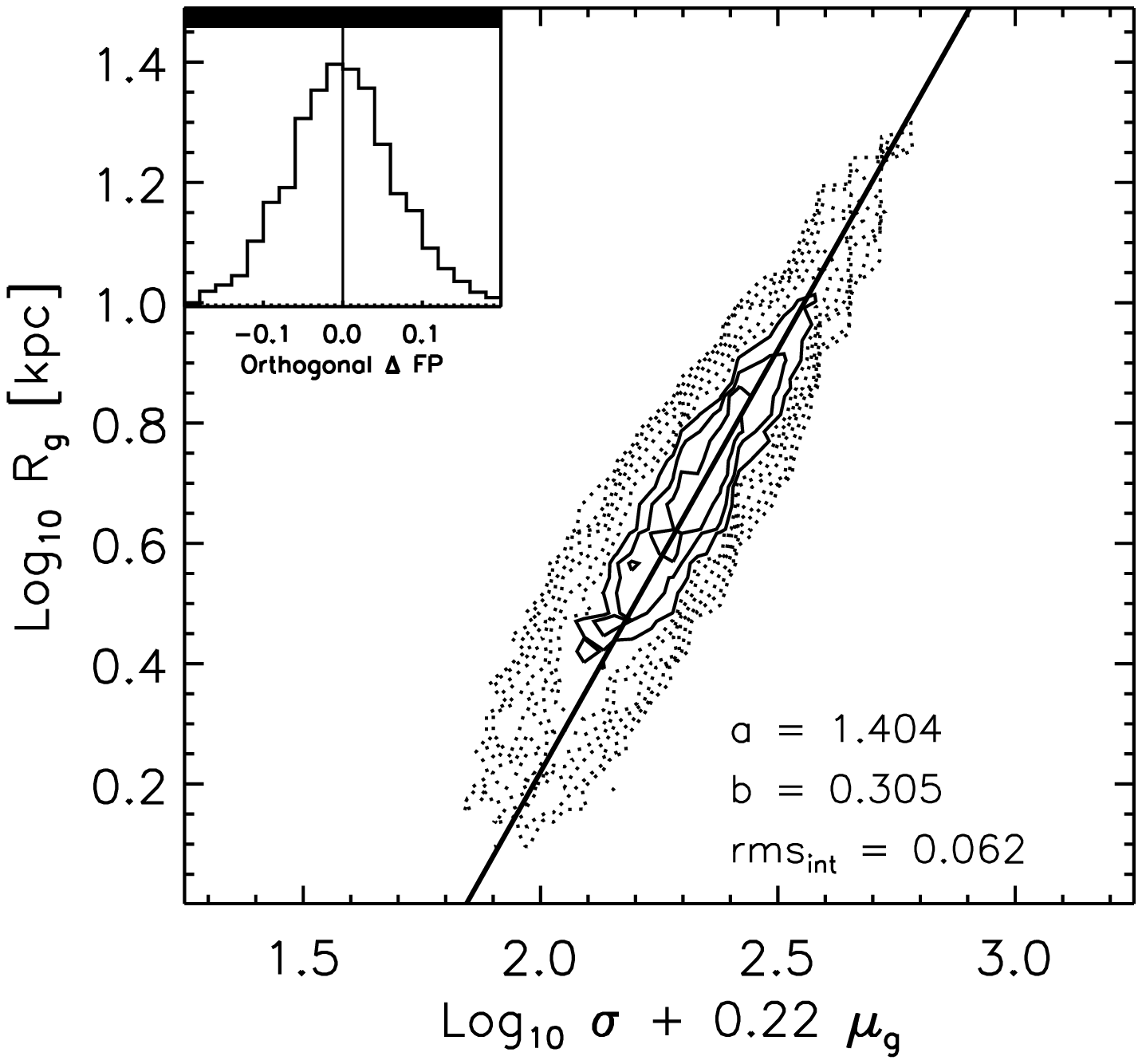}
 \includegraphics[width=0.475\hsize]{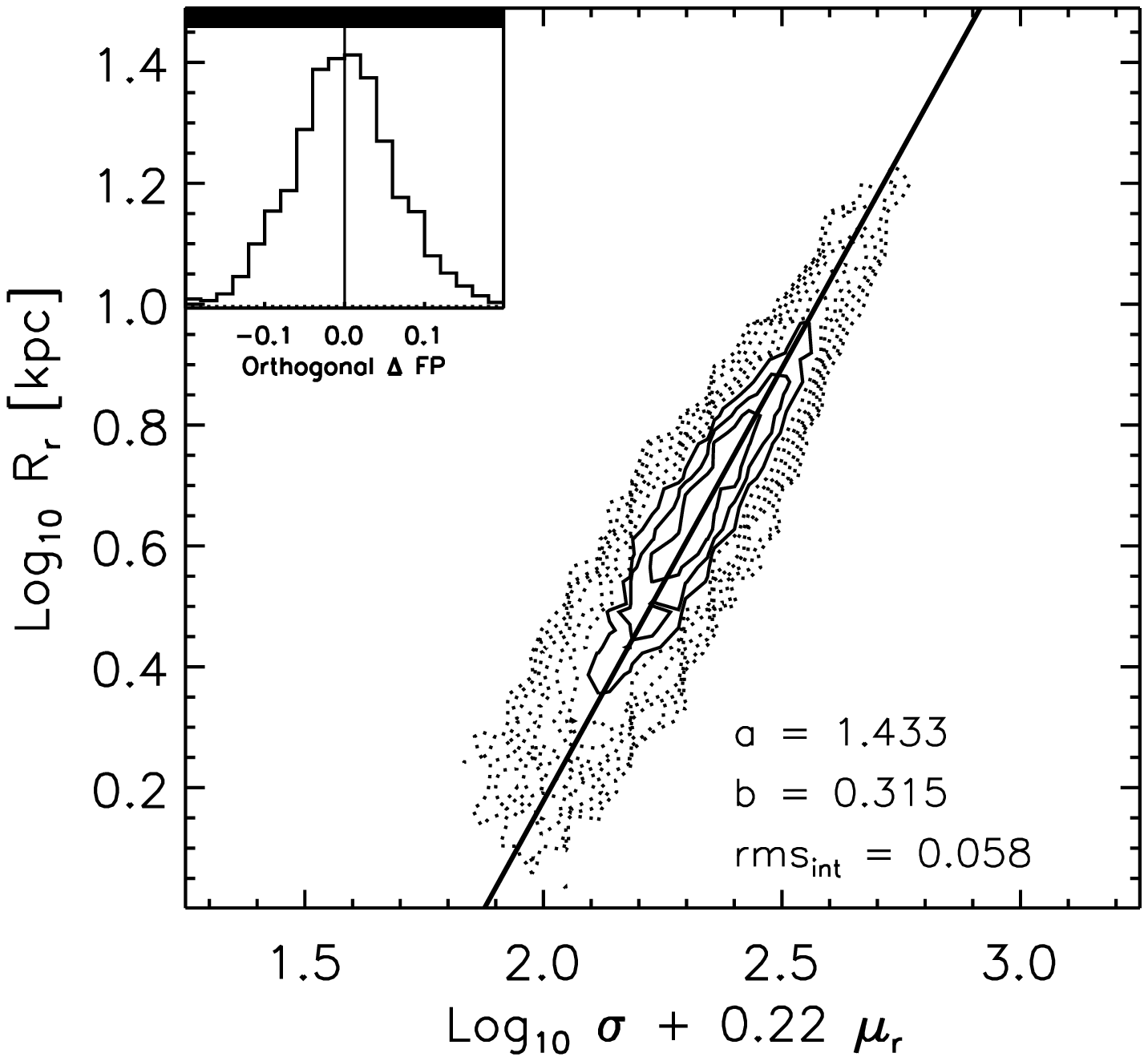}
 \includegraphics[width=0.475\hsize]{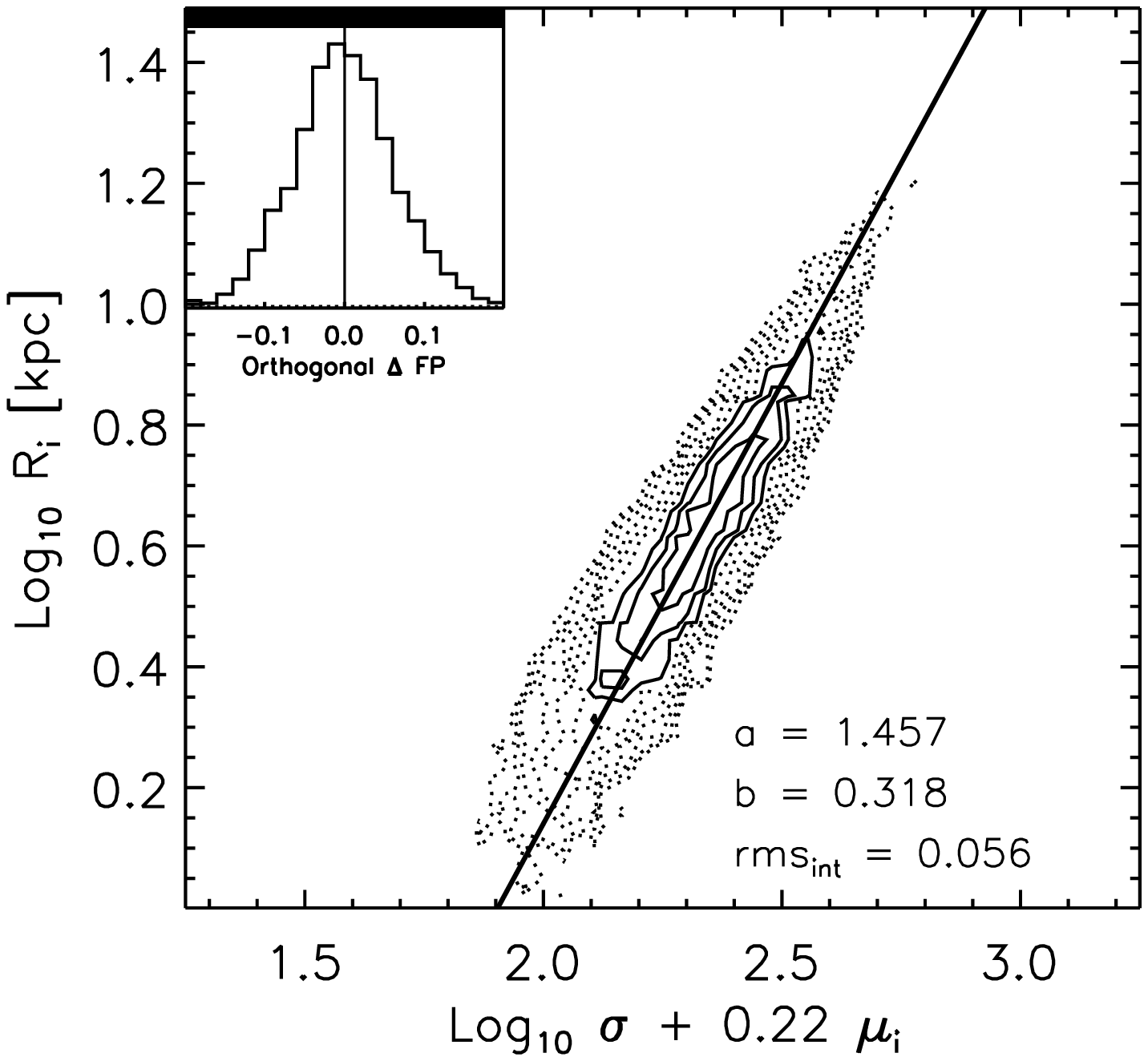}
 \includegraphics[width=0.475\hsize]{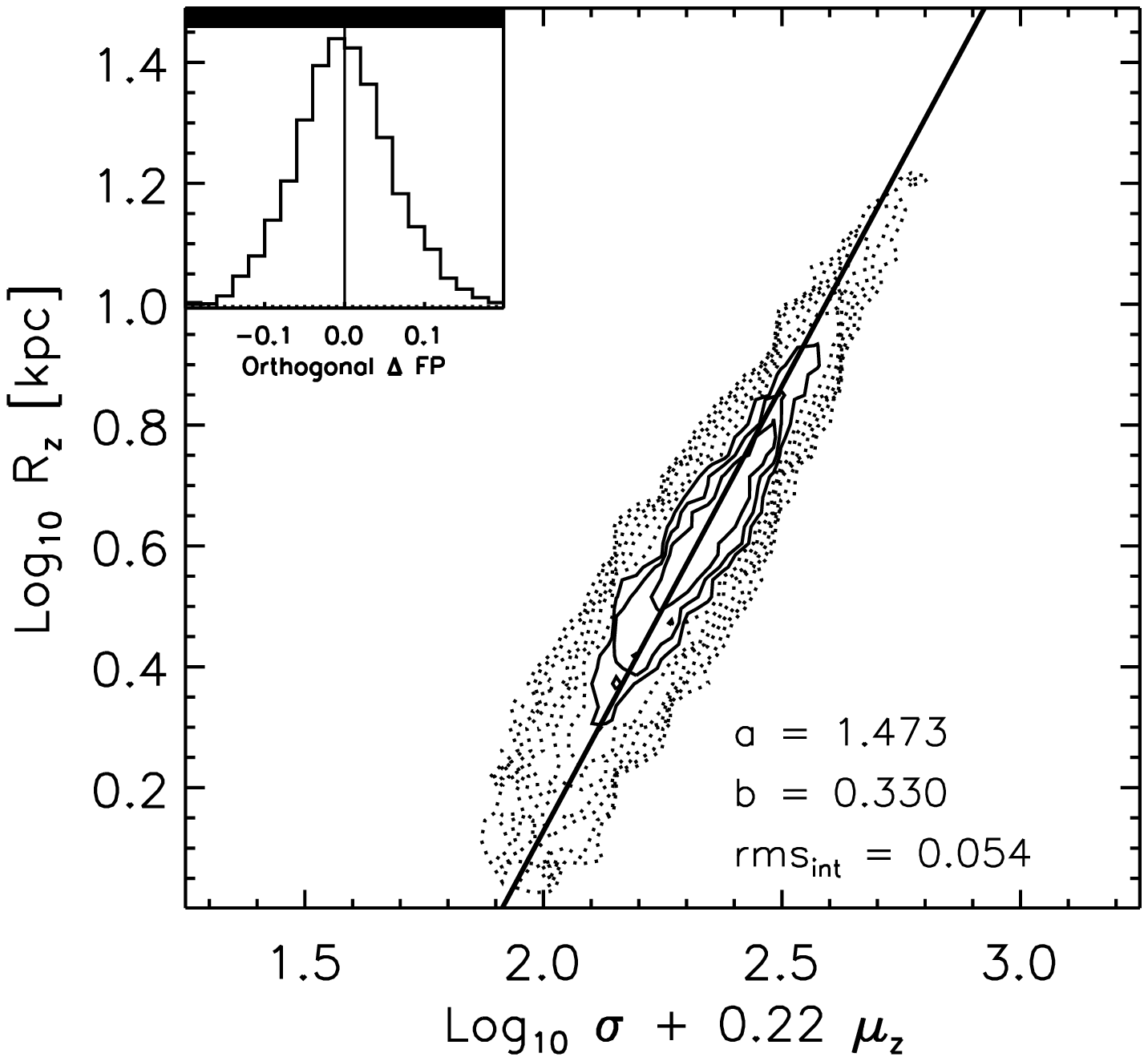}
 \caption{The Fundamental Plane in the $g$, $r$, $i$ and $z$ bands.
Typical uncertainties on the coefficients due to random errors 
are $\delta a \sim 0.02$ and $\delta b \sim 0.01$. Sytematics errors 
give $\delta a \sim 0.05$ and $\delta b \sim 0.02$.}
 \label{FPgriz}
\end{figure*}

\begin{figure*}
 \centering
 \includegraphics[width=0.475\hsize]{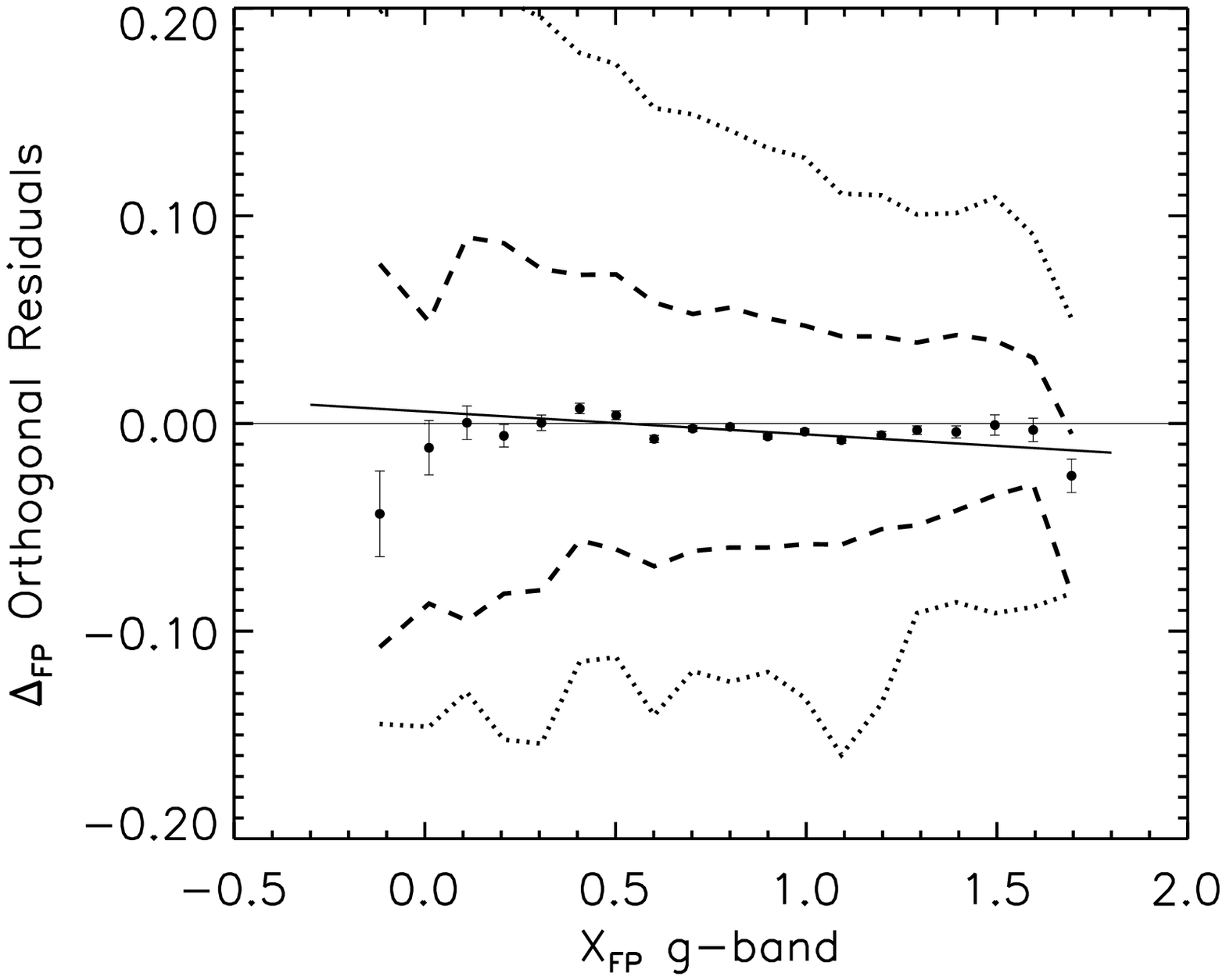}
 \includegraphics[width=0.475\hsize]{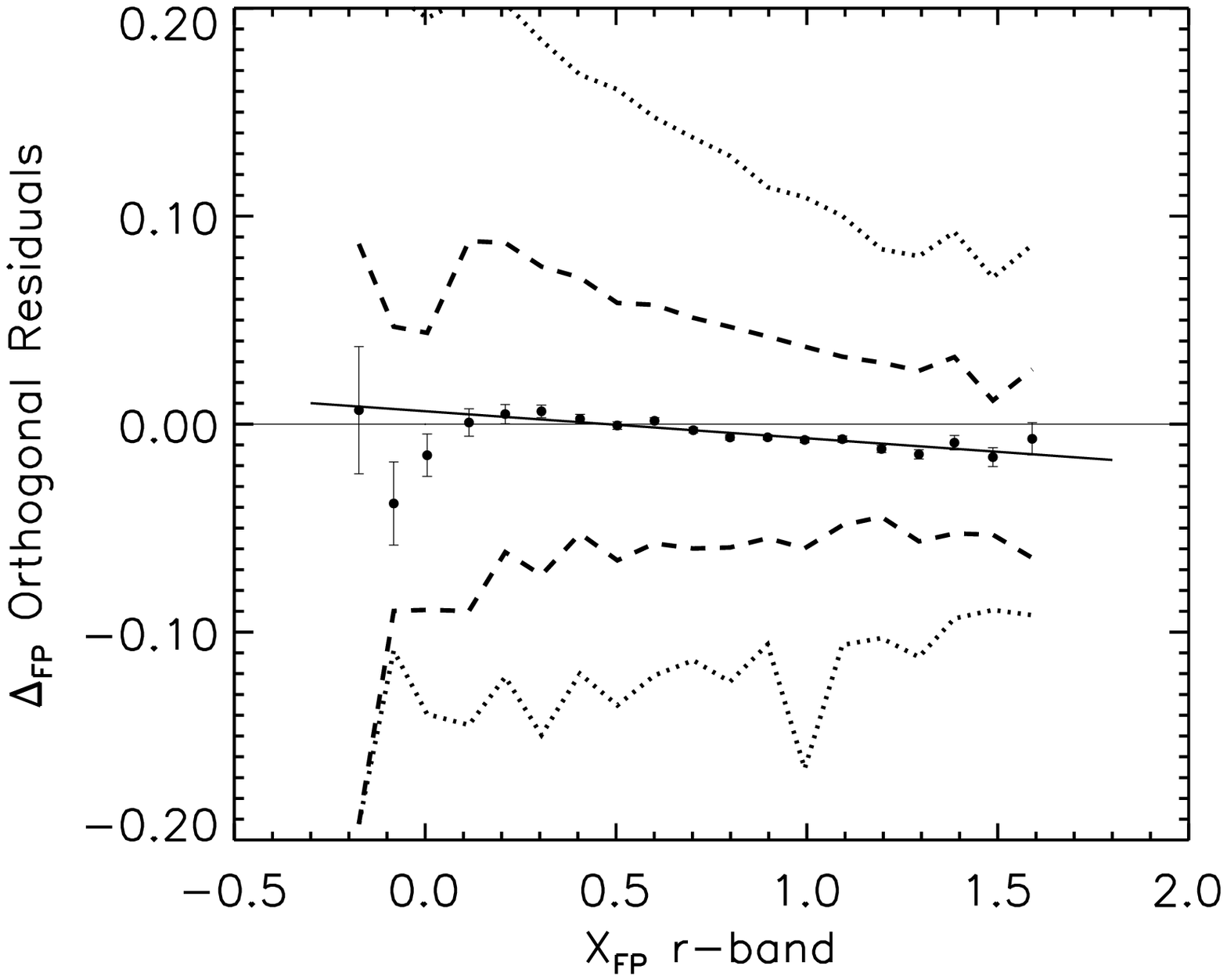}
 \includegraphics[width=0.475\hsize]{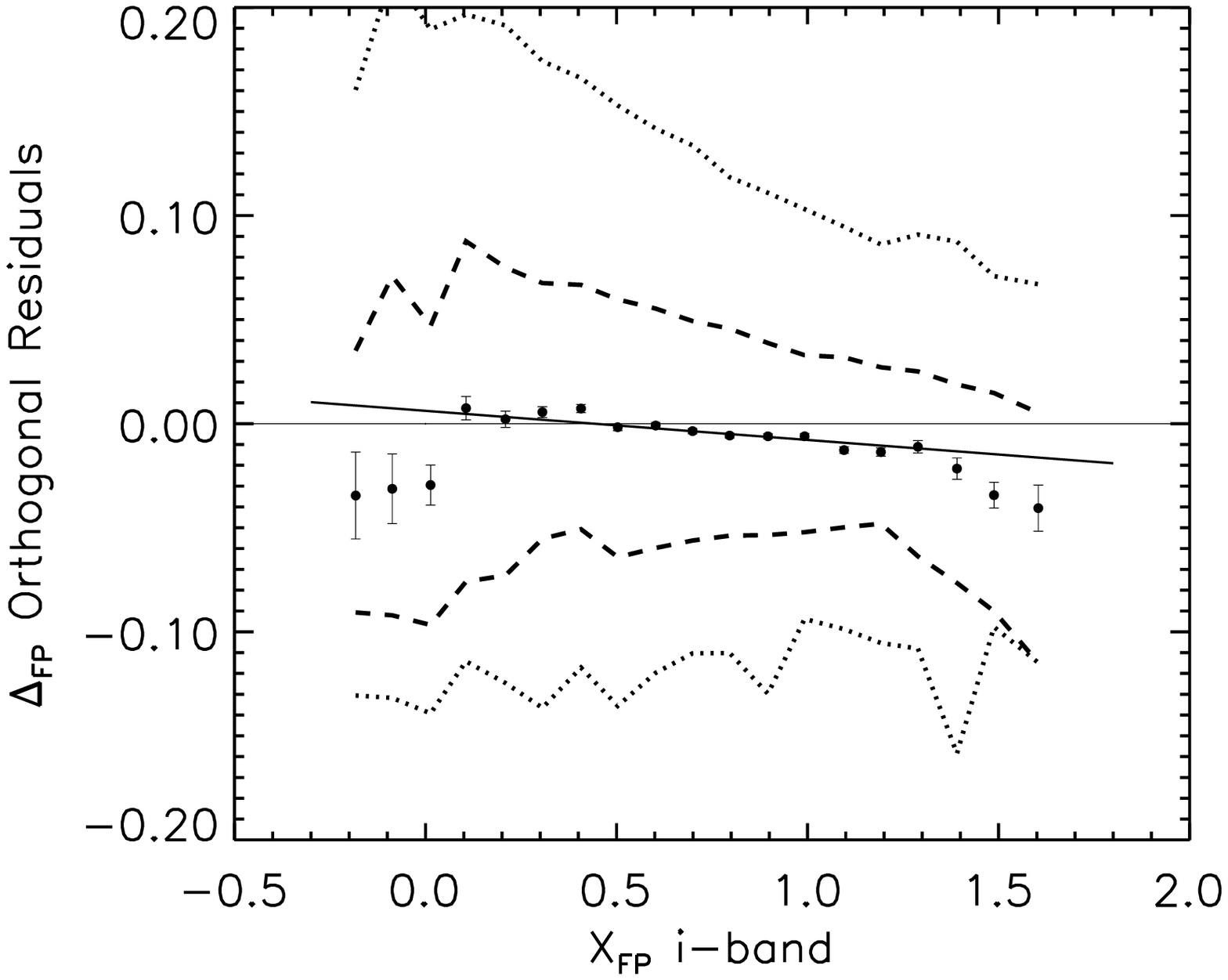}
 \includegraphics[width=0.475\hsize]{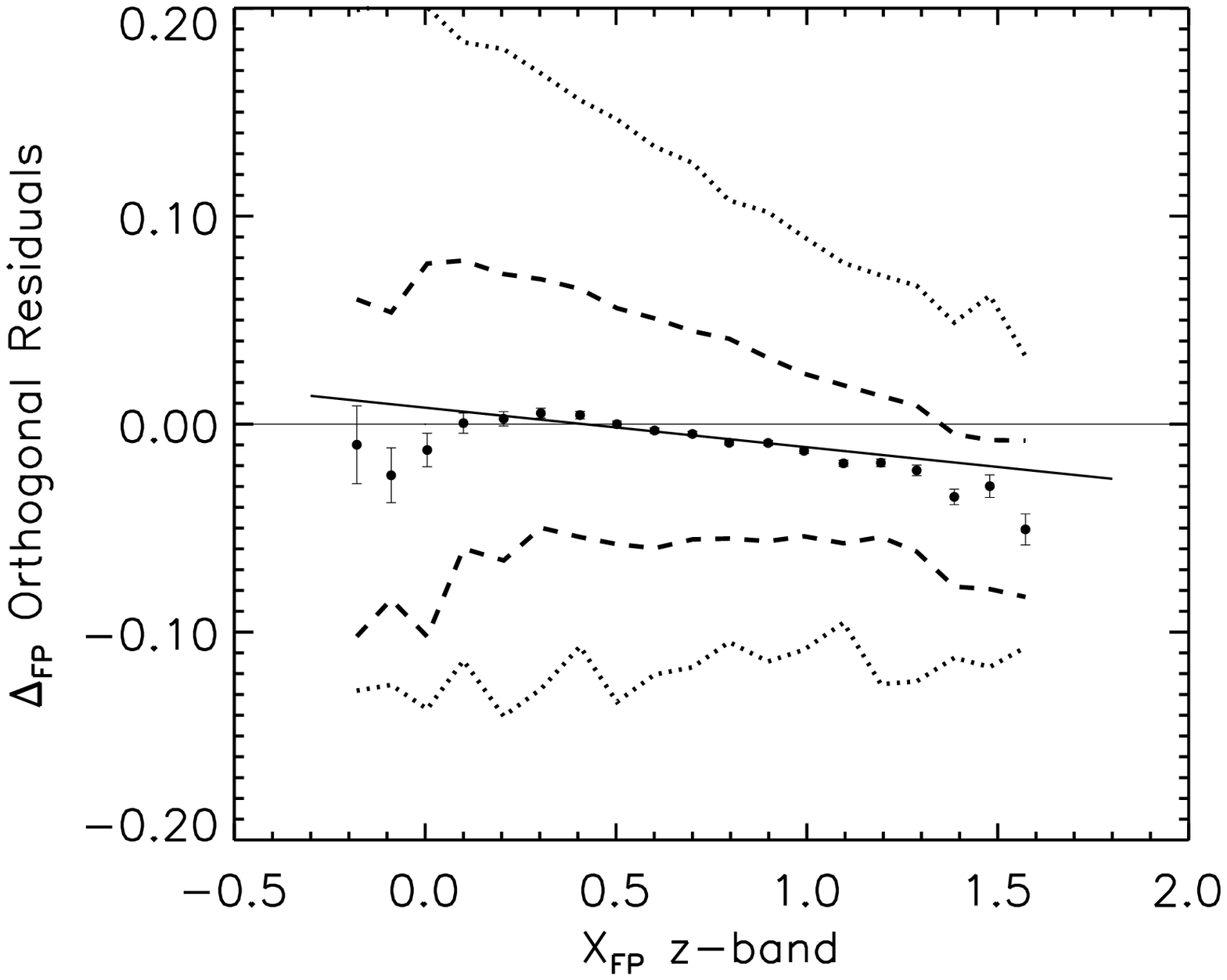}
 \caption{Fundamental Plane residuals with respect to the orthogonal 
          fit in the $g$, $r$, $i$ and $z$ bands, shown as a function 
          of distance X$_{\rm FP}$ along the Plane.  Symbols show the 
          median residual in narrow bins in $X_{\rm FP}$; dashed 
          and dotted lines enclose 68\% and 95\% of the points.  Solid 
          lines show the expected trend due to correlated measurement 
          errors. }
 \label{residFPgriz}
\end{figure*}

When working with a large galaxy sample, systematic effects may be 
more important than random errors. 
The correction applied to the magnitudes, sizes and stellar masses 
for known problems which arise from the SDSS sky-subtraction algorithm 
(Hyde \& Bernardi 2009), results in small changes to $a$ and $b$ --
the variation is less than $0.01$ for both coefficients.

However, recall that our estimate of the velocity dispersion is 
the average of the DR6 and {\tt IDLspec2d} values.  
If we only use the DR6 values, we find 
$( a_{\rm direct-DR6}, b_{\rm direct-DR6}) = (1.189,0.303)$ 
with intrinsic scatter $0.108$~dex, and 
 $( a_{\rm orth-DR6}, b_{\rm orth-DR6}) = (1.464,0.315)$ 
with scatter $0.058$~dex.  
Using the {\tt IDLspec2d} reductions instead gives 
$( a_{\rm direct-IDLspec2d}, b_{\rm direct-IDLspec2d}) = (1.122,0.304)$, with intrinsic scatter $0.100$~dex, and 
 $( a_{\rm orth-IDLspec2d},\bar b_{\rm orth-IDLspec2d}) = (1.401,0.317)$ 
with scatter $0.061$~dex, respectively.
This suggests that we should assume as typical systematic error 
$\delta a \sim 0.05$ and $\delta b \sim 0.02$.  
These values are larger than the random errors quoted above, showing 
that is important to separate random from systematic errors.

\subsection{Dependence on selection}

If we do not account for selection effects we find mean values 
 $\langle M_r \rangle = -21.94$, $\langle\log R_e\rangle = 0.62$, 
 $\langle\mu_e \rangle = 19.71$ and $\langle\log \sigma \rangle = 2.30$, 
and covariances 
$ C_{II} = 0.2660$, $ C_{RR} = 0.0488$, $ C_{VV} = 0.0127$, 
$ C_{IR} = 0.0820$, $ C_{IV} = 0.0036$ and $ C_{RV} = 0.0159$.
Although these covariances have changed, the coefficients of the 
Fundamental Plane change little:  
 $(a_{\rm direct},b_{\rm direct}) = (1.1723,0.2924)$ 
with intrinsic scatter $0.079$, and 
 $(a_{\rm orth},b_{\rm orth}) = (1.4235,0.2914)$ 
with scatter $0.047$.  Note that the intrinsic scatter is 
smaller!  This is largely because the distribution of luminosities 
in a magnitude limited survey is narrower than in a volume limited 
survey (because the faint objects are under-represented); if one 
does not account for this, then the magnitude limited catalog is 
more homogeneous, making for a tighter FP.  
The fact that the coefficients $a$ and $b$ hardly change 
is largely fortuitous (Sheth \& Bernardi 2009 show why they are 
expected to change); it happens because the scaling relations are 
actually slightly curved (see Section~\ref{curved}, and also 
Hyde \& Bernardi 2009).  

Had we not removed objects with $b/a<0.6$ then 
 $C_{II} = 0.3756$, $C_{RR} = 0.0549$, $C_{VV} = 0.0171$, 
 $C_{IR} = 0.0948$, $C_{IV} = -0.0132$ and $C_{RV} = 0.0171$.  
The resulting FP has 
 $(a_{\rm direct},b_{\rm direct}) = (1.1674,0.2936)$ 
with intrinsic scatter $0.091$, and 
 $(a_{\rm orth},b_{\rm orth}) = (1.4294,0.3052)$ with scatter $0.055$.  
Since objects with $b/a<0.6$ are almost certainly not early-types, 
the fact that $a$ and $b$ are almost the same as when these objects 
have been removed suggests that the presence of a few non-early-types 
in our sample has little effect on our findings.

Finally, selecting only galaxies with SDSS-DR6 velocity dispersions 
(i.e. excluding objects with low $S/N$ spectra or presence of weak 
emission lines, i.e. the {\tt status} flag not-equal to 4) changes 
$a$ and $b$ very slightly  -- the variation is less than $0.01$ for 
both coefficients.  Table~\ref{otherFPs} compares the $r$-band 
FP coefficients associated with these various selection and parameter 
choices.  

\begin{table}
  \begin{center}
   \caption{Dependence of $r$-band Fundamental Plane coefficients on sample selection and parameters.  Typical uncertainties on the coefficients due to random errors are $\delta a \sim 0.02$ and $\delta b \sim 0.01$; sytematics are $\delta a \sim 0.05$ and $\delta b \sim 0.02$.}\label{otherFPs}
    \begin{tabular}{c c c c c}
      \hline
      Band & $a$ & $b$ & rms$_{int}$ \\
      \hline
      direct  &     &            &    \\
      $r$     &1.1701 & 0.3029 & 0.0964\\
      DR6-$\sigma$ &1.1892 & 0.3032 & 0.1081\\
      spec2d-$\sigma$ &1.1223 & 0.3041 & 0.1002 \\
      no-Vmax &1.1703 & 0.3036 & 0.0792\\
      all $b/a$ & 1.1674 & 0.2936 & 0.0913 \\
      \hline
      orthog &     &        &            &       \\
      $r$     &1.4335 & 0.3150 & 0.0578\\
      DR6$\sigma$ &1.4642 & 0.3151 & 0.0581 \\
      spec2d$\sigma$ &1.4013 & 0.3173 & 0.0613 \\
      no-Vmax &1.4235 & 0.2914 & 0.0473\\
      all $b/a$ & 1.4294 & 0.3052 & 0.0554\\
      \hline
    \end{tabular}
  \end{center}
\end{table}

\subsection{Dependence on waveband}

The Fundamental Planes associated with the orthogonal fits to the 
$g$, $r$, $i$ and $z$ band data are shown in Figure~\ref{FPgriz}.  
The coefficients $(a,b)$ of these Fundamental Planes are shown 
in each panel; they are also reported in Table~\ref{griz}, as are 
the corresponding coefficients of the direct fits.  Notice that 
there is a small but systematic increase of $a$ with wavelength.  
To estimate its significance, we require an estimate of the errors 
on $a$ and $b$.  
Although systematics associated with how $\sigma$ was measured make 
$\delta a \sim 0.05$, $\delta b \sim 0.02$, this additional systematic error 
is not relevant if we wish to compare the Planes at different 
wavelengths, because the same choice for $\sigma$ is made for all 
wavelengths.  What matters here is the typical uncertainty due to 
random errors on these best fit values.  
Section~\ref{errors} and Figure~\ref{err_coef} shows that these 
random errors are $\delta a \sim 0.02$, $\delta b \sim 0.01$.  
Thus, we have a $3\sigma$ detection of the steepening of the 
Fundamental Plane with wavelength.

\begin{table}
  \begin{center}
   \caption{Coefficients $(a,b)$ of the Luminosity Fundamental Plane. 
Typical uncertainties on the coefficients due to random errors 
are $\delta a \sim 0.02$ and $\delta b \sim 0.01$. Sytematics errors 
give $\delta a \sim 0.05$ and $\delta b \sim 0.02$.}\label{griz}
    \begin{tabular}{c c c c c c}
      \hline
      Band & $a$ & $b$ & $c$ & rms$_{obs}$ & rms$_{int}$ \\
      \hline
      direct &     &        &         &        &       \\
      $g$  &1.1154 & 0.2957 & -8.0463 & 0.1102 & 0.1005\\
      $r$  &1.1701 & 0.3029 & -8.0858 & 0.1074 & 0.0964\\
      $i$  &1.1990 & 0.3036 & -8.0481 & 0.1067 & 0.0950\\
      $z$  &1.2340 & 0.3139 & -8.2161 & 0.1052 & 0.0921\\
      \hline
      orthog &     &        &         &        &       \\
      $g$  &1.4043 & 0.3045 & -8.8579 & 0.0696 & 0.0617\\
      $r$  &1.4335 & 0.3150 & -8.8979 & 0.0664 & 0.0578\\
      $i$  &1.4572 & 0.3182 & -8.8914 & 0.0652 & 0.0563\\
      $z$  &1.4735 & 0.3295 & -9.0323 & 0.0635 & 0.0538\\
      \hline
    \end{tabular}
  \end{center}
\end{table}

\subsection{Evidence for curvature}\label{curved}
Figure~\ref{residFPgriz} shows residuals 
\begin{equation}
 \Delta_{\rm FP}\equiv \frac{\log R_e - a\log\sigma - b \mu_e - c}
                            {\sqrt{1 + a^2 + b^2}}
\end{equation} 
from the orthogonal fit as a function of distance 
\begin{equation}
 X_{\rm FP} \equiv \frac{a\log R_e + \log \sigma + b\,\mu_e/a}
                        {\sqrt{1 + a^2}}
\end{equation}
along the Plane.  Weak trends are seen in all bands.

Interpretation of these trends is complicated by the fact that 
the measurement errors are correlated, and they are larger at 
small sizes.  To see what effect is expected, we must use the 
numbers from the covariance matrices ${\cal F}$ and ${\cal E}$ to 
estimate 
 $\langle \Delta_{\rm FP}X_{\rm FP}\rangle/\langle X_{\rm FP}^2\rangle$, 
the expected slope of the correlation shown in Figure~\ref{residFPgriz}.  
In the $r$-band, 
 $\langle \Delta_{\rm FP}X_{\rm FP}\rangle = 0 - 0.0007$, 
where the first term is the contribution from ${\cal F}$ and the 
second from ${\cal E}$.  I.e., there is no intrinsic correlation 
(essentially, by definition), but there is a contribution from 
the measurement errors.  This contribution is easy to estimate 
because $\Delta_{\rm FP}$ is almost proportional to IP $- aV$, 
where IP $\equiv \log_{10} ( R_e ) - 0.3\mu_e$, and the error 
in IP is well-known to be negligible, at least for deVaucouleur-like 
profiles (Saglia et al. 1997; or see bottom panel of Figure~5 in 
Hyde \& Bernardi 2009), so the error in $\Delta_{\rm FP}$ is 
dominated by the error in $V$.  
On the other hand, the error in $X_{\rm FP}$ is due to errors in 
$R$, $V$ and IP.  Errors in $R$ and $V$ are almost uncorrelated, 
so the dominant contribution from correlated errors comes from the 
fact that $V$ appears with different signs in $X_{\rm FP}$ and 
$\Delta_{\rm FP}$.  

A similar analysis of $\langle X_{\rm FP}^2\rangle$ 
shows that it is dominated by the contribution from ${\cal F}$ 
rather than ${\cal E}$.  In the $r$-band, 
$\langle X_{\rm FP}^2\rangle = 0.087 + 0.008$.  
Thus, correlated errors are expected to produce a weak trend, 
with slope $-0.007$.  The solid line in the Figure shows this 
trend -- it is similar to that observed, suggesting that correlated 
errors can account for most of the observed weak decline in 
$\Delta_{\rm FP}$.  
On the other hand, correlated errors cannot account for the break 
downwards at small $X_{\rm FP}$:  this is genuine curvature.  

A similar analysis in the other bands yields the solid lines 
shown in Figure~\ref{residFPgriz}.  
This illustrates that sample sizes are now large enough that 
correlated errors produce systematic effects which must be 
accounted for when performing the fits.  

At the low size/mass end, the Plane is probably warped 
(Figures~\ref{residFPgriz}, and see also Figure~\ref{FPmstars} below).
However, we view this with caution since uncertainties/systematics in 
the velocity dispersion/size measurements are also larger. The possibility
of curvature induced by the presence of spirals in the sample is discussed
in Appendix A. We argue that the location of spirals on the fundamental Plane
would not result in negative residuals at small $X_{\rm FP}$ , and that the spirals
 are to few to significantly bias our results.

\subsection{Variation in thickness along the Plane}
\begin{figure}
 \centering
 \includegraphics[width=0.85\hsize]{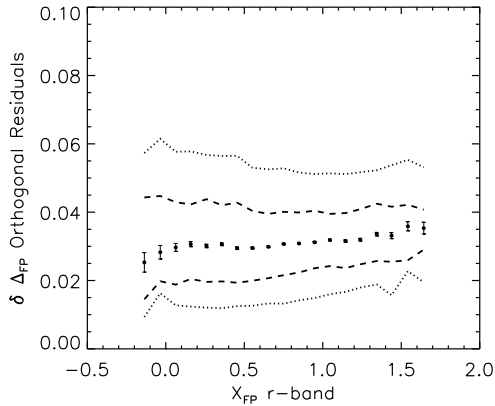}
 \caption{Contribution of measurement errors to the residuals from 
          the orthogonal fit as a function of $X_{\rm FP}$ for the 
          $r$-band (symbols and line styles same as previous figure).  
          Results for other bands are similar.  }
 \label{FPerr}
\end{figure}

Notice that the Plane is thinner at large $X_{\rm FP}$.
Figure~\ref{FPerr} shows that the typical measurement errors vary 
little with distance along the Plane.  Since the observed scatter 
is larger at smaller $X_{\rm FP}$ (the width of the regions 
enclosed by the dashed and dotted curves in Figure~\ref{residFPgriz} 
increases at small $X_{\rm FP}$), but the contribution from errors 
is approximately constant, we conclude that the intrinsic scatter 
around the Plane decreases dramatically as $X_{\rm FP}$ increases.  
It is remarkable that, at $X_{\rm FP}\ga 1$, measurement errors 
account for essentially all of the observed scatter (about 0.04~dex), 
suggesting that the Plane is rather thin at large $X_{\rm FP}$.

\subsection{Potential biases from cuts in $L$}\label{cutl}
There has been recent interest in the fact that the coefficients 
of the FP depend on how the sample was selected, a point 
emphasized by Bernardi et al. (2003).  E.g., 
D'Onofrio et al. (2008) and Nigoche-Netro et al. (2008) show that 
$a$ decreases if faint galaxies are removed from the sample, and 
one does not account for the fact that they are missing.  
Donofrio et al. suggest that this may reflect the fact that 
luminous and faint galaxies had different formation histories.  
We show below that, although the latter may be true, the dependence 
of $a$ on sample selection alone is not a reliable indicator.  
We do so by constructing an FP relation using pairwise correlations 
with {\em no} curvature -- so one would have concluded that nothing 
special was happening to galaxies at either end of the sequence.  
We then show that cuts in $L$ change $a$ and $b$ in ways that are 
quite similar to that observed by D'Onofrio et al.  
In the following subsection, we show that similar effects also 
occur if cuts in $\sigma$ are made.  

Before discussing the FP, it is easier to first consider the 
size-surface brightness relation.  At fixed $L$, 
$\langle\log_{10}R|\mu_e\rangle$ will be a line of slope $1/5$ 
with no scatter, {\em by definition}.  Changing $L$ moves this 
line to the left or right, but does not change the slope.  
However, because there is a correlation between $R_e$ and $L$, 
more luminous galaxies are bigger on average, the high $L$ galaxies 
only populate the large $R$ part of their line; galaxies of lower 
$L$ only populate the lower part of their line.  
Thus, in general, the full $\log R-\mu_e$ correlation, which is 
got by averaging over the full range in $L$, will have a different 
slope than $1/5$, with the difference being determined by the 
strength of the $\log R_e - \log L$ correlation.  

\begin{figure}
 \centering
 \includegraphics[width=\hsize]{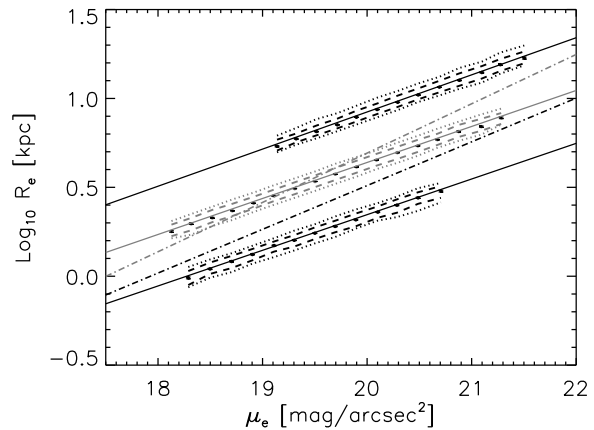}
 \caption{Effective radius versus surface-brightness in magnitude bins of
          (from botton to top) $-20.5 < M_r < -20$, $-22 < M_r < -21.5$ and
          $-23.5 < M_r < -23$. The solid lines are linear fit to the 
          galaxies in the
          different bins. The slope is $0.2$ in all cases.  Dashed and 
          dotted lines show the regions which contain 68\% and 95\% of 
          the objects in each bin.  Upper dot-dashed curve shows a 
          fit to the objects in all three bins, and lower dot-dashed 
          curve shows the correlation in the full data-set which 
          accounts for selection effect (as in Table~1 of Hyde \& Bernardi 2009).}
 \label{muRtest}
\end{figure}

Figure~\ref{muRtest} shows this explicitly using three narrow 
non-overlapping bins in $L$.  In any one bin, the slope of the 
relation (shown by the solid lines) is $0.2$.  The slope obtained 
from combining two nearby bins will be slightly steeper; combining 
all three bins would yield an even steeper slope.
This is a generic argument, but note that it also works when 
$\langle \log R_e|\log L\rangle$ increases linearly with $\log L$.
Since the slope of the $\log R_e-\mu_e$ correlation is $1/5$ 
for a narrow bin in $L$, but something else when all $L$ are 
included, the slope of the $\log R_e - \mu$ correlation will 
appear to depend on the range of $L$ included in the sample 
{\em even though the fundamental underlying correlation is linear}.  
This, essentially, is the origin of the behaviour seen by 
D'Onofrio et al. (2008) and Nigoche-Netro et al. (2008). 

To show this explicitly, the upper of the two dot-dashed lines in 
the Figure shows the result of averaging over the objects in the 
three luminosity bins shown in Figure~\ref{muRtest}:  it is steeper 
than the relation for any one of the bins.  Of course, in a magnitude 
limited survey, one does not simply average over all the luminosity 
bins.  Rather, when objects with the full range of $L$ are used, then 
each object is weighted by $V_{\rm max}^{-1}(L)$.  The lower dot-dashed 
line shows this relation:  this shows that both the slope and the 
zero-point of this correlation are sensitive to this weighting.

Extending this argument to the Fundamental Plane is slightly 
more involved.  Once again, at fixed $L$, the scaling between 
$R$ and $\mu$ is straightforward.  The inclusion of a 
$\sigma$-dependent term to the x-axis serves to shift the lines 
associated with different $\sigma$ horizontally -- the goal of 
the FP algorithm is to shift them so they lie on top of one another 
as much as possible.  
To see the effect this has, note that for a given $L$, each object 
with size $R_e$ is shifted by an amount which depends on its 
$\sigma$, so the full Fundamental Plane consists of shifting each 
point horizontally in the $\log R_e-\mu_e$ Plane by 
 $\langle \log\sigma|\log L,\log R_e\rangle$ 
on average.  If the mean shift depends on $R_e$, this will change 
the slope of the line of fixed $L$ from $1/5$ to something else.  
If there is scatter around this mean shift, then the line of 
fixed $L$ will be broadened.  The combination of change in slope 
and additional scatter both serve to make the parameters $a$ 
and $b$ of the Fundamental Plane depend on the range of $L$ which 
are included in the fit, even though the underlying pairwise 
correlations are linear (i.e., their slope does not depend on $L$).  

\begin{figure}
 \centering
 \includegraphics[width=\hsize]{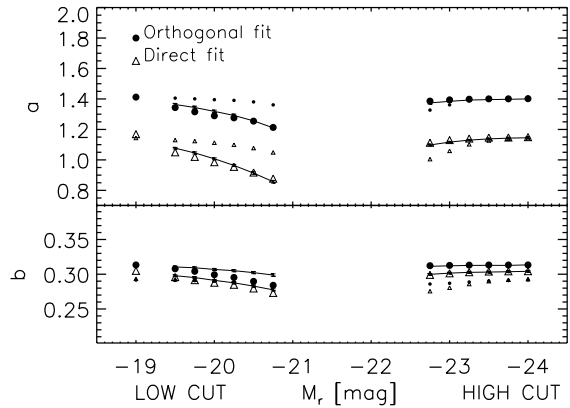}
 \caption{Dependence of FP coefficients on the luminosity range of 
          the sample.  The set of symbols on the left show how 
          $a$ and $b$ change if low luminosity objects are excluded 
          from the sample (the symbols show the absolute magnitude 
          of the faintest object which is kept in the sample); 
          the symbols on the right show what happens if high 
          luminosity objects are excluded (symbols show the 
          absolute magnitude of the brightest object in the sample).  
          Large symbols show the result of accounting for the 
          selection effect which comes from the apparent magnitude 
          limit of the SDSS; smaller symbols show the result of 
          ignoring this effect.  Smooth curves show similar 
          measurements in a mock catalog in which all underlying 
          correlations were pure power-laws.}
 \label{FPabL}
\end{figure}

Figure~\ref{FPabL} shows all this explicitly.  The symbols show 
how the FP coefficients change as faint objects are excluded 
(symbols on the left) or as luminous objects are excluded (symbols 
on the right).  Large symbols show the effect of accounting for 
selection effects (each object is weighted by the inverse of the
volume over which it could have been observed, so luminous objects 
have smaller weights), and smaller symbols show results when all 
objects are weighted equally (selection effects are ignored).  
Note that $a$ depends strongly on the luminosity cut, whereas $b$ 
is less strongly affected.  
The smooth curves show the result of making similar measurements 
in a mock catalog constructed following methods given in 
Bernardi et al. (2003).  Note in particular that all scaling 
relations in the mock catalog were linear; there was no curvature.  
Nevertheless, Figure~\ref{FPabL} shows that the FP coefficients in 
the mock catalog depend on the value of the luminosity threshold 
similarly to how they do in the data.  This demonstrates that a 
detection of dependence of $a$ on luminosity threshold does 
{\em not} imply that the underlying scaling relations are curved.

The strong dependence of $a$ on luminosity threshold is easily 
understood.  At fixed $L$, $R$ and $\sigma$ are anti-correlated; 
it is only when averaged over a large range in $L$ that $R$ and 
$\sigma$ are positively correlated (e.g. Bernardi et al. 2003).  
Since $a$ is essentially the slope of the $R-\sigma$ relation, 
the result of restricting the range in $L$ is to drive $a$ to 
smaller values, since at fixed $L$ it must be negative.  This 
also explains why, when no account is taken of the magnitude 
limited selection, $a$ changes little (compared to when the 
selection effect is accounted for) when low luminosity 
galaxies are excluded, but more strongly when high luminosities 
are excluded.  In a magnitude limited sample, the luminosity 
distribution is biased to larger $L$ (since they can be seen 
to larger volumes), so the effect of removing low $L$ is less 
severe (this removes a smaller fraction of galaxies), but the 
effect of a cut at high $L$ is more dramatic (this removes a 
larger fraction of the galaxies).  

\begin{figure}
 \centering
 \includegraphics[width=\hsize]{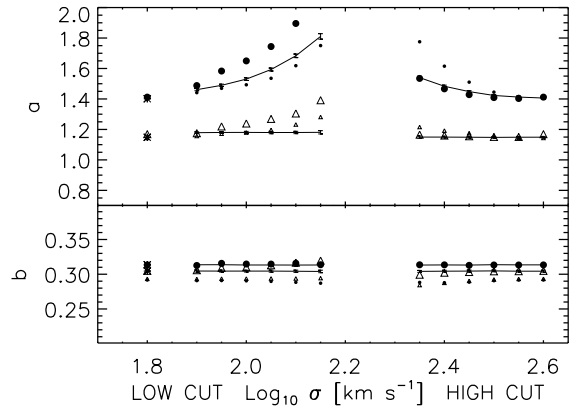}
 \caption{Dependence of FP coefficients on the range of velocity 
          dispersions in the sample.  The set of symbols on the 
          left show measurements in samples in which objects with 
          $\sigma$ less than the value indicated were excluded; 
          the set of symbols on the right show results when the 
          sample excludes objects with $\sigma$ larger than the 
          indicated value.  Large symbols show the result of 
          accounting for the selection effect which comes from 
          the apparent magnitude limit of the SDSS; smaller symbols 
          show the result of ignoring this effect.  Smooth curves 
          show similar measurements in a mock catalog in which all 
          underlying correlations were pure power-laws.}
 \label{FPabV}
\end{figure}

\subsection{Potential biases from cuts in $\sigma$}\label{cutv}
There is no fundamental reason why we should have restricted our 
study to how $a$ and $b$ depend on $L$.  This subsection shows how 
the FP coefficients change as the range of $\sigma$ in the sample 
is varied.  This study is potentially more interesting than that 
of the previous subsection, since few samples are selected on the 
basis of absolute magnitude, but many samples exclude objects 
with small $\sigma$, simply because small velocity dispersions are 
difficult to measure.  

Figure~\ref{FPabV} shows that $b$ is (again) hardly affected.  
However, $a_{\rm orth}$ depends strongly on this selection, and the 
trend is opposite to that for luminosity, while $a_{\rm direct}$ is 
less affected (also see Bernardi et al. 2003; Nigoche-Netro et al. 2008).  
This is easy to understand (also see discussion in Bernardi et al. 2003):  
$a_{\rm direct}$ is close to the slope $C_{RV}/C_{VV}$ of the 
$\langle \log R|\log\sigma\rangle$ relation.  If this relation is 
linear, then excluding small or large values of $\sigma$ should not 
change this slope.  On the other hand, $a_{\rm orth}$ measures the 
orthogonal slope of this relation, and this must steepen as the bin 
in $\sigma$ narrows; after all, in the limit of a very narrow bin in 
$\sigma$, this orthogonal slope must become infinite.  
(Sheth \& Bernardi 2009 show this analytically -- $a_{\rm orth}$ 
depends both on $C_{RV}/C_{VV}$ and on $C_{VV}$ itself.  Reducing 
the range of $\sigma$ reduces $C_{VV}$, thus increasing $a_{\rm orth}$.)  
Once again, similar measurements in the pure power-law mock produce 
similar trends, although in this case the agreement with the SDSS is 
not as good.  
Most of this (small) discrepancy can be attributed to the fact that 
the underlying pairwise scaling relations in the SDSS are slightly 
curved (see next section), an effect which we have deliberately 
removed from our mocks so as to illustrate that Figures~\ref{FPabL} 
and~\ref{FPabV} are not good diagnostics of curvature.

\subsection{Comparison with previous work}
Our fits differ slightly from those reported by 
Bernardi et al. (2003).  This is not unexpected, because the SDSS 
improved its photometric reductions significantly between DR1 (on 
which Bernardi et al. 2003 was based) and DR6 (on which the present 
analysis is based).  In addition, Figure~\ref{FPabV} shows that 
simply removing $\sigma < 100$~km~s$^{-1}$ increases $a_{\rm orth}$ 
by about 15\%.  
This is interesting because Bernardi et al. (2003) removed objects 
with $\sigma<90$~km~s$^{-1}$ from their sample (due to the 
dispersion of the SDSS spectrograph).  
The distribution of velocity dispersions is relatively narrow, 
so this cut can have a non-negligible effect.  
Figure~\ref{FPabV} shows that this might bias $a$ high by about 
10\%, which is about the level of discrepancy between the Bernardi 
et al. (2003) FP and this work (but note that the maximum likelihood 
method of Bernardi et al. did try to account for the cut in $\sigma$.)  

While the absolute values of the best-fit coefficients have changed, 
the relative values have not changed significantly:  both $a$ and 
$b$ increase while the intrinsic scatter decreases in the redder 
bands.  This is true whether the fit minimizes the scatter in the 
$\log R_e$ direction, or in the direction orthogonal to the Plane.  
In addition, it is thinner at the large $\log R_e$ end.  
The weak but significant increase of $a$ with wavelength is 
consistent with that reported by Bernardi et al. (2003) and, more 
recently, by La Barbera et al. (2008).

\begin{figure}
 \centering
 \includegraphics[width=\hsize]{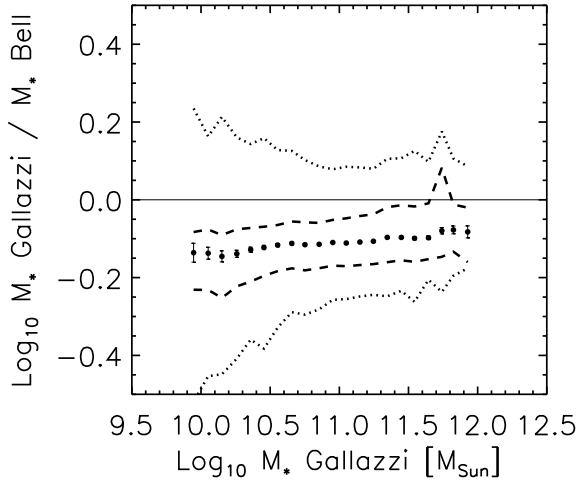}
 \caption{Comparison of the stellar mass estimates from spectra 
          (Gallazzi et al. 2005) with those from broad-band photometry 
          (Bell et al. 2003).  Symbols and line-styles same as in 
          Figure~\ref{residFPgriz}.}
 \label{checkMstars}
\end{figure}

\begin{figure*}
 \centering
 \includegraphics[width=0.475\hsize]{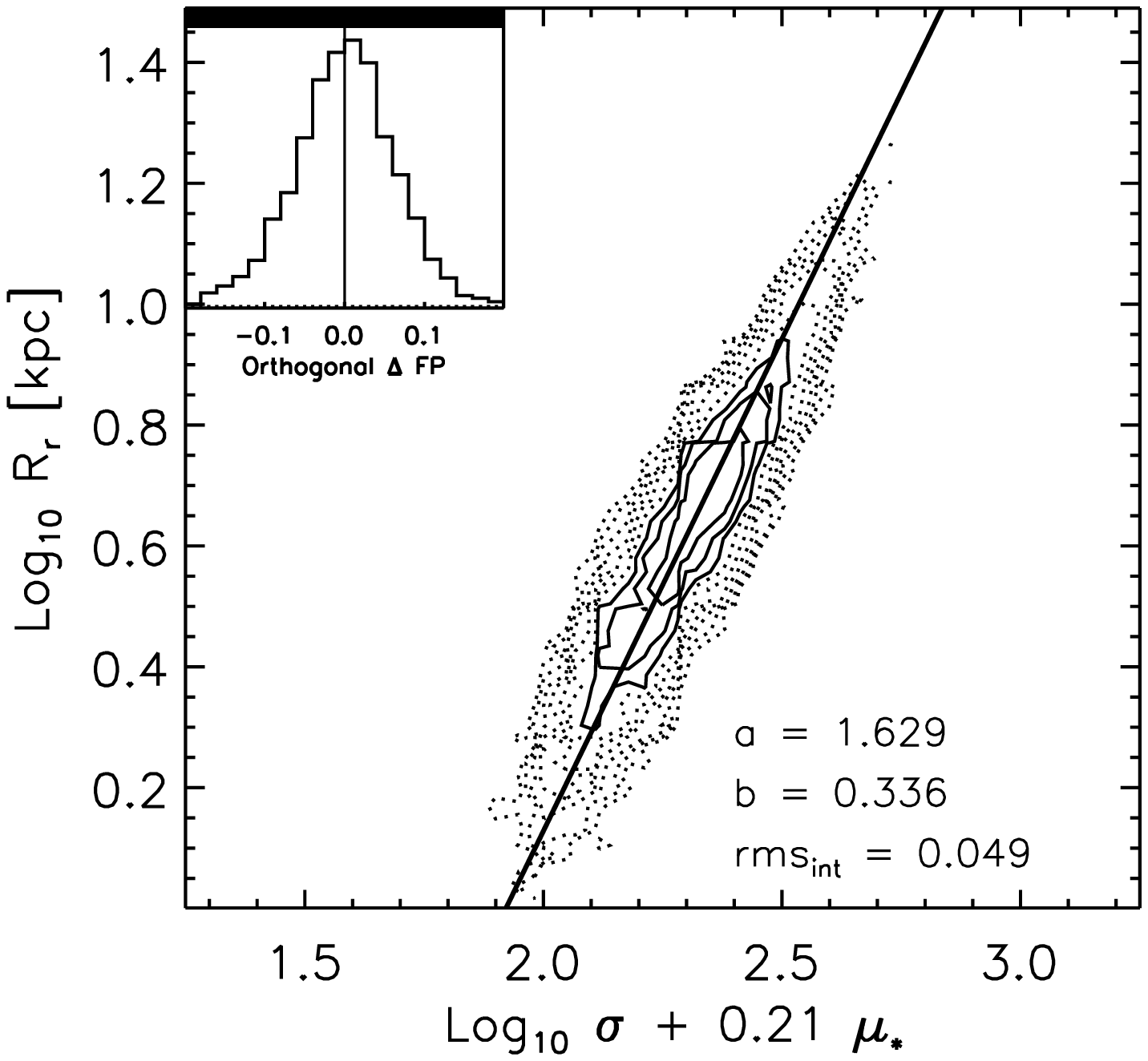}
 \includegraphics[width=0.475\hsize]{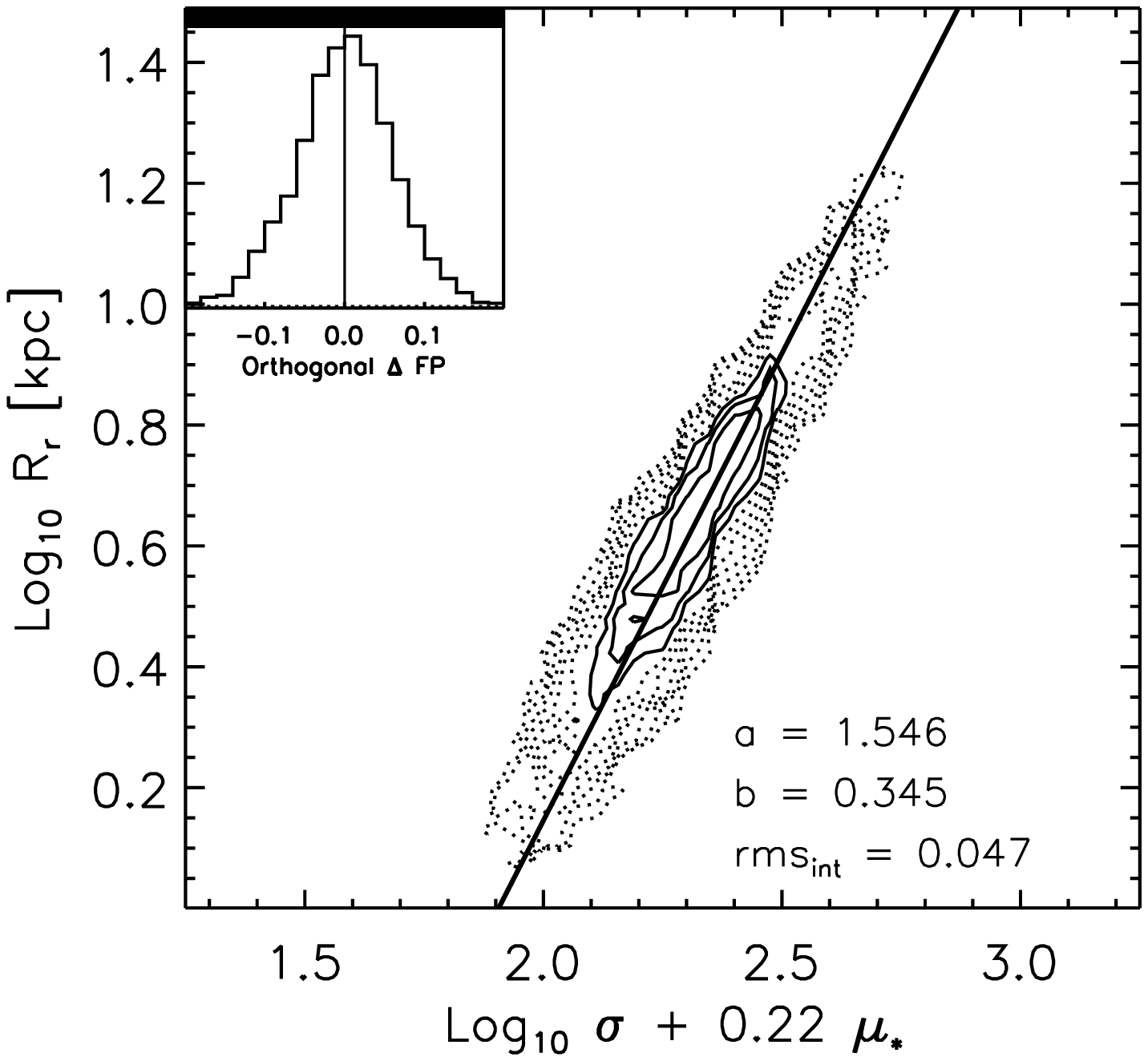}
 \includegraphics[width=0.475\hsize]{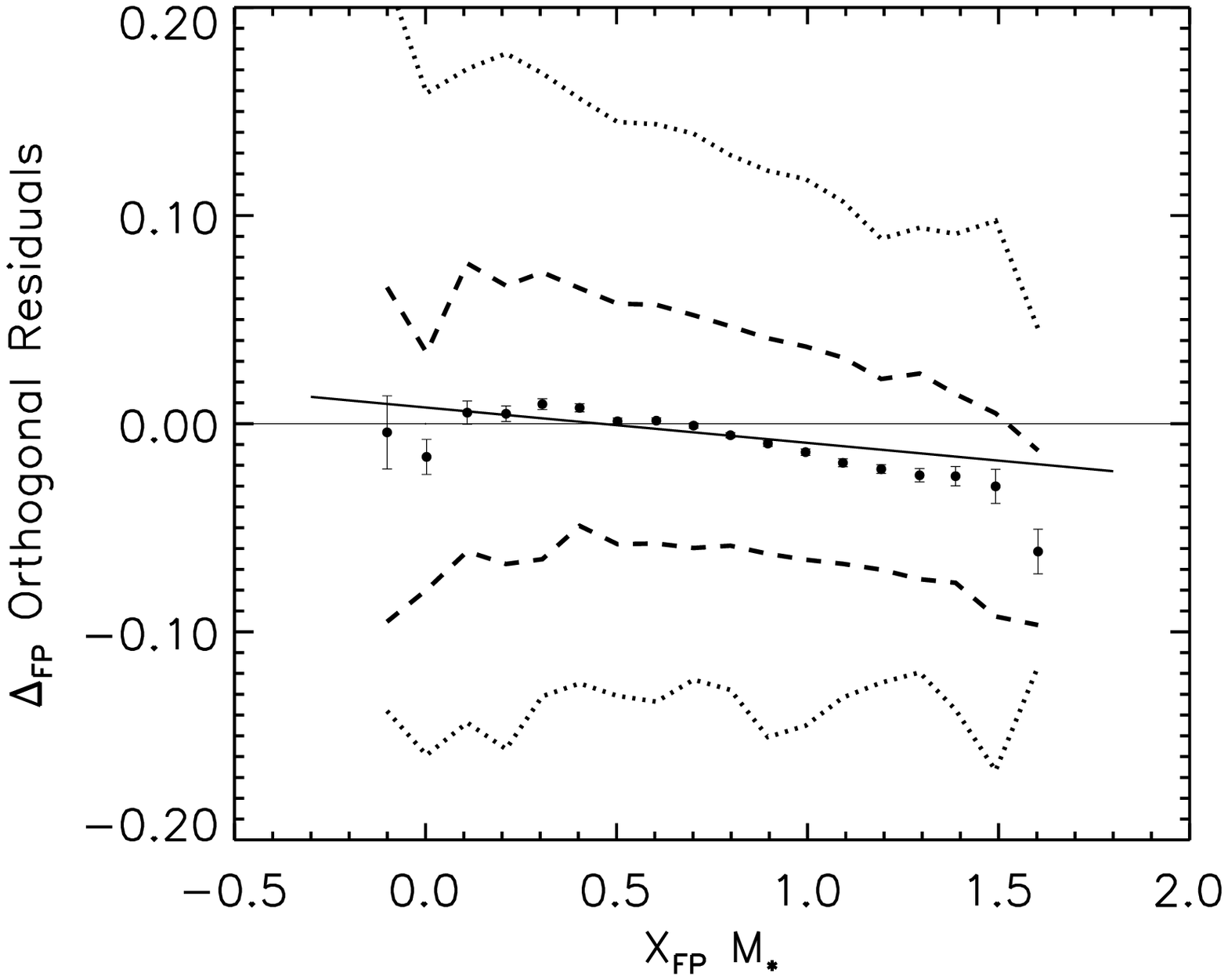}
 \includegraphics[width=0.475\hsize]{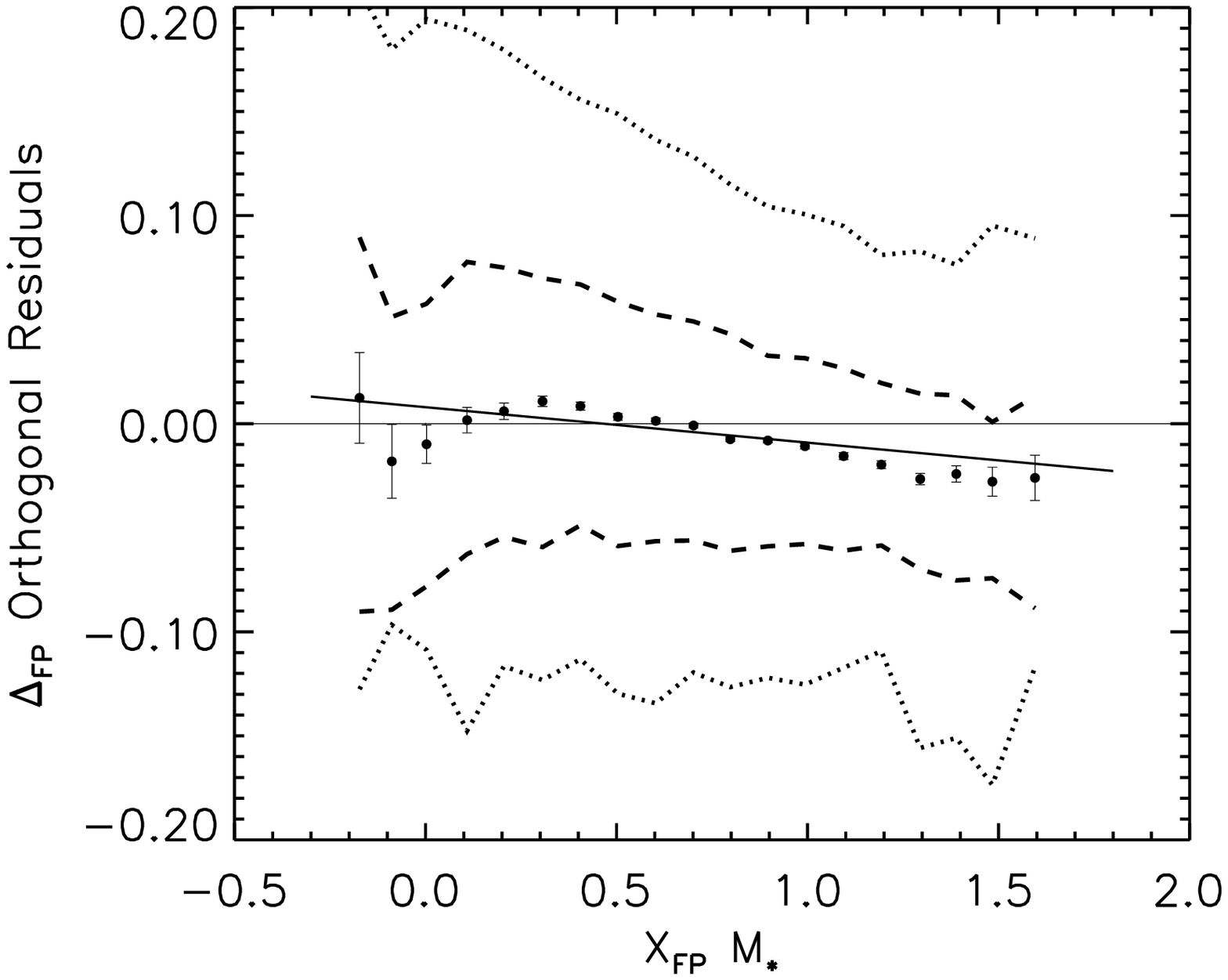}
 \caption{The stellar mass Fundamental Plane associated with 
          the two esimates of the stellar mass
          (from spectra, left and photometry, right).  Symbols 
          and line-styles same as for Figures~\ref{FPgriz} 
          and~\ref{residFPgriz}. Typical uncertainties on the coefficients 
due to random errors 
are $\delta a \sim 0.02$ and $\delta b \sim 0.01$. Sytematics errors 
give $\delta a \sim 0.05$ and $\delta b \sim 0.02$.}
 \label{FPmstars}
\end{figure*}

\section{FP$_*$:  The Stellar Mass Fundamental Plane}\label{FPmass}
The previous section studied the traditional Fundamental Plane, FP.  
In this section, we show the result of replacing luminosities 
with stellar masses to produce what we will call FP$_*$.  
To illustrate that our results are robust to changes in how 
one estimates the stellar mass, we show results based on two 
different estimates of $M_*/L$:  
one, from Gallazzi et al. (2005), which is based on a likelihood 
analysis of the spectra; 
another, from Bell et al. (2003), who use the $k$-corrected 
broad-band color
 $(M_*/L_r) = 1.097\,(g-r) - 0.306$.  
Figure~\ref{checkMstars} compares these estimates with each 
other; there is an offset of about 0.15~dex and a small trend 
which shows that $M_{\rm Gallazzi}/M_{\rm Bell}$ tends to increase
at larger stellar mass. Most of this offset (0.11~dex) is due to 
the difference in initial mass function (IMF) used in the stellar 
population models by the two groups. (Gallazzi et al. used 
Bruzual \& Charlot 2003 models which assume the Chabrier 2003 IMF, 
while Bell et al. assume a diet-Saltpeter IMF.)  The rms scatter 
between these two estimates increases significantly at small $M_*$.  
While the offset will affect the zero-point but not the slope of 
the FP$_*$, the weak trend could actually introduce a small 
systematic effect in the slope.

Figure~\ref{FPmstars} shows the associated Fundamental Planes, 
the coefficients of which are reported in Table~\ref{abMst}.  
The uncertainties in $\delta a$ and $\delta b$ are similar to those
described in Section~\ref{errors} (i.e., random errors $\delta a \sim 0.02$ 
and $\delta b \sim 0.01$, with larger systematic uncertainties of
 $\delta a \sim 0.05$ and $\delta b \sim 0.02$).
In this case also, understanding the errors is important.  
We have assumed that the error in $\mu_*$ is given by the error 
in the photometric quantity $\mu$ plus a contribution from the 
error in $\log M_*/L$.  Gallazzi et al. (2005) report rms errors 
of about 0.06~dex, so we add 2.5 times 0.06~dex in quadrature.  
We further assume that the error in $\log M_*/L$ is uncorrelated 
with that in the size or velocity dispersion estimates.  
Whereas the first assumption is probably correct, the second is 
almost certainly incorrect for the stellar mass estimates which 
come from the spectra.  (Gallazzi et al. 2005 do not show how 
the error on $M_*/L$ correlates with the error on other observables.
They do show that the $M_*/L$ error does not correlate with spectral type.)  
This may explain some of the differences we see between the two 
stellar mass Planes.

The important point is that in both cases, the slope of FP$_*$ is 
steeper than that of the FP in the $z$-band, but it is shallower 
than the virial scaling.  This suggests that the ratio of 
dynamical to stellar  mass, $M_{dyn}/M_*$, varies systematically across the 
population.  Our results suggest that it is a weakly increasing 
function of $M_{dyn}$ or $M_*$.  Figure~\ref{showMs} shows a direct 
comparison:
 $\langle M_{dyn}/M_*|M_{dyn}\rangle \propto M_{dyn}^{0.17\pm 0.01}$ 
where $M_{dyn} \equiv 5 R_e\sigma^2/G$ and $M_*$ is from Gallazzi et al. 
(the error 0.01 on the slope 
was computed accounting for systematics errors -- the uncertainty from 
random errors is smaller $\sim 0.003$). 
If there were no scatter around this relation, then we would expect 
 $\langle M_{dyn}/M_*|M_*\rangle \propto M_*^{0.17/0.83}\propto M_*^{0.2}$; 
because there is scatter, this scaling is shallower, 
$\propto M_*^{0.06\pm0.01}$ (Hyde \& Bernardi 2009).  
Gallazzi et al. (2006) also measured the $M_{dyn}$, $M_*$ relationship
using deVaucouleur radii of SDSS galaxies. They obtain  $\langle M_{dyn}/M_*|M_{dyn}\rangle \propto M_{dyn}^{0.19\pm 0.03}$ 
which is statistically equivalent to our result.

\section{The Plane in $\kappa$-space}\label{kspace}

\begin{table}
  \begin{center}
   \caption{Coefficients $(\alpha,\beta)$ of the Stellar Mass 
            Fundamental Plane. 
Typical uncertainties on the coefficients due to random errors 
are $\delta a \sim 0.02$ and $\delta b \sim 0.01$. Sytematics errors 
give $\delta a \sim 0.05$ and $\delta b \sim 0.02$.}\label{abMst}
    \begin{tabular}{c c c c c c}
      \hline
      $M_*/L$ & $\alpha$ & $\beta$ & c & rms$_{obs}$ & rms$_{int}$\\
      \hline
      direct   & & & & & \\
      Spectra  & 1.3989 & 0.3164 & 4.4858 & 0.1160 & 0.0894\\
      Color    & 1.3501 & 0.3293 & 5.0015 & 0.1115 & 0.0835\\
      \hline
      orthog   & & & & & \\
      Spectra  & 1.6287 & 0.3359 & 4.4238 & 0.0648 & 0.0486\\
      Color    & 1.5462 & 0.3449 & 4.9300 & 0.0638 & 0.0466\\

      \hline
    \end{tabular}
  \end{center}
\end{table}

\begin{figure}
 \centering
 \includegraphics[width=\hsize]{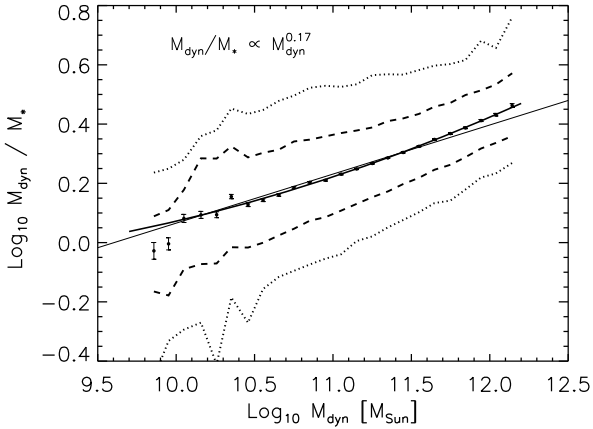}
 \caption{Ratio of dynamical and stellar masses (from Gallazzi et al. 2005)
          as a function of 
          dynamical mass.  Symbols and line-styles 
          same as Figure~\ref{residFPgriz}. The thick solid curve is a quadratic fit (see Hyde \& Bernardi 2009).}
 \label{showMs}
\end{figure}

Bender, Burstein \& Faber (1992) suggested that the combination of 
variables which makes up the Fundamental Plane could be combined 
in a more physically transparent way.  They called this combination 
$\kappa$-space.  The three axes of this space are 
\begin{eqnarray}
 \kappa_1 &\equiv& \frac{\log_{10}(R_e\sigma^2)}{\sqrt{2}}, \qquad
 \kappa_2  \equiv  \frac{\log_{10}(\sigma^2I_e^2/R_e)}{\sqrt{6}}, \quad 
         {\rm and}\nonumber\\
 \kappa_3 &\equiv& \frac{\log_{10}(R_e\sigma^2/L)}{\sqrt{3}}.
\end{eqnarray}
It is interesting to examine how early types are distributed in 
what we call $\kappa_*$-space, which is obtained by replacing 
luminosity with (Gallazzi et al. 2005) stellar mass in the expressions above.

\begin{figure*}
 \centering
 \includegraphics[width=0.475\hsize]{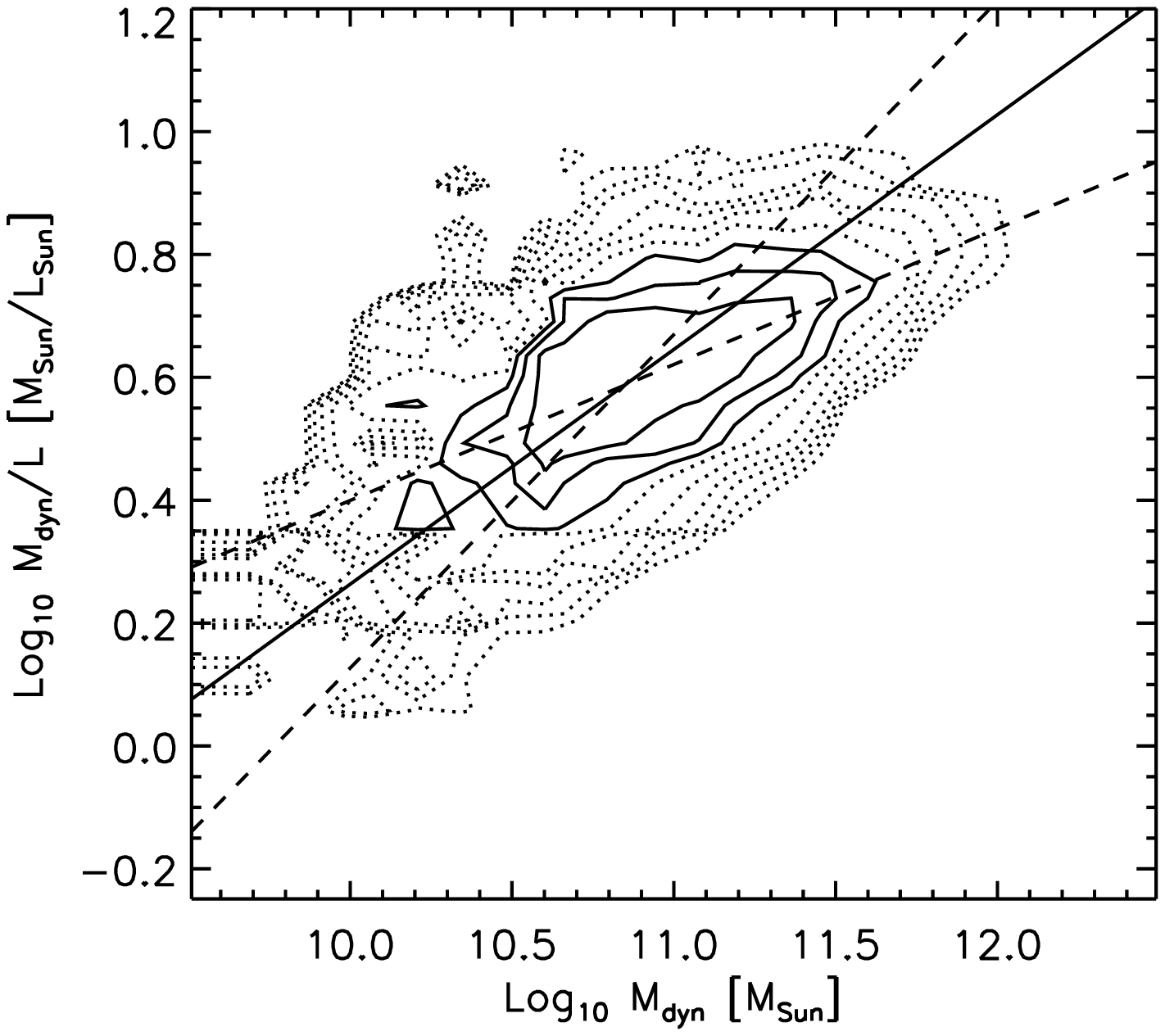}
 \includegraphics[width=0.475\hsize]{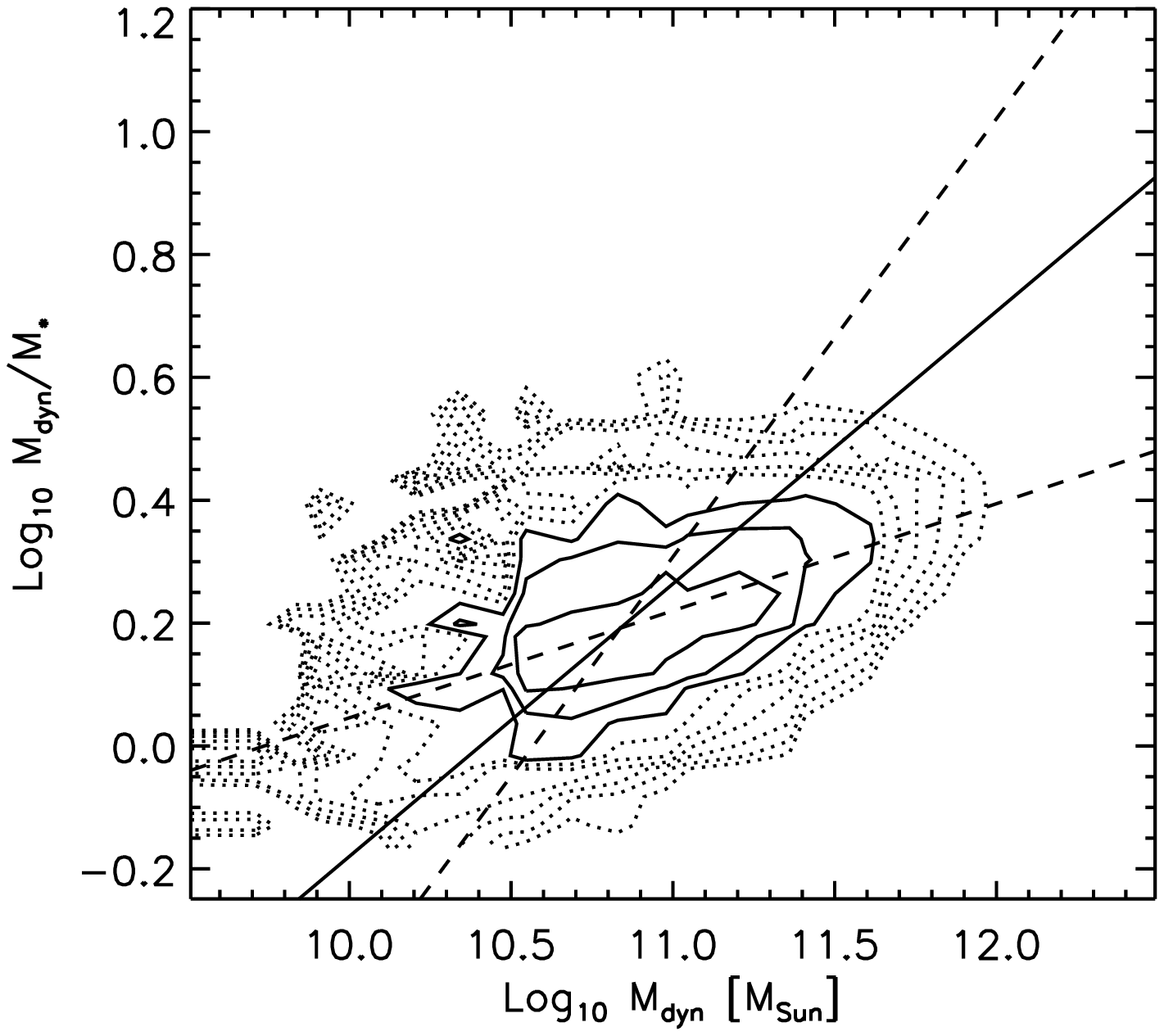}
 \includegraphics[width=0.475\hsize]{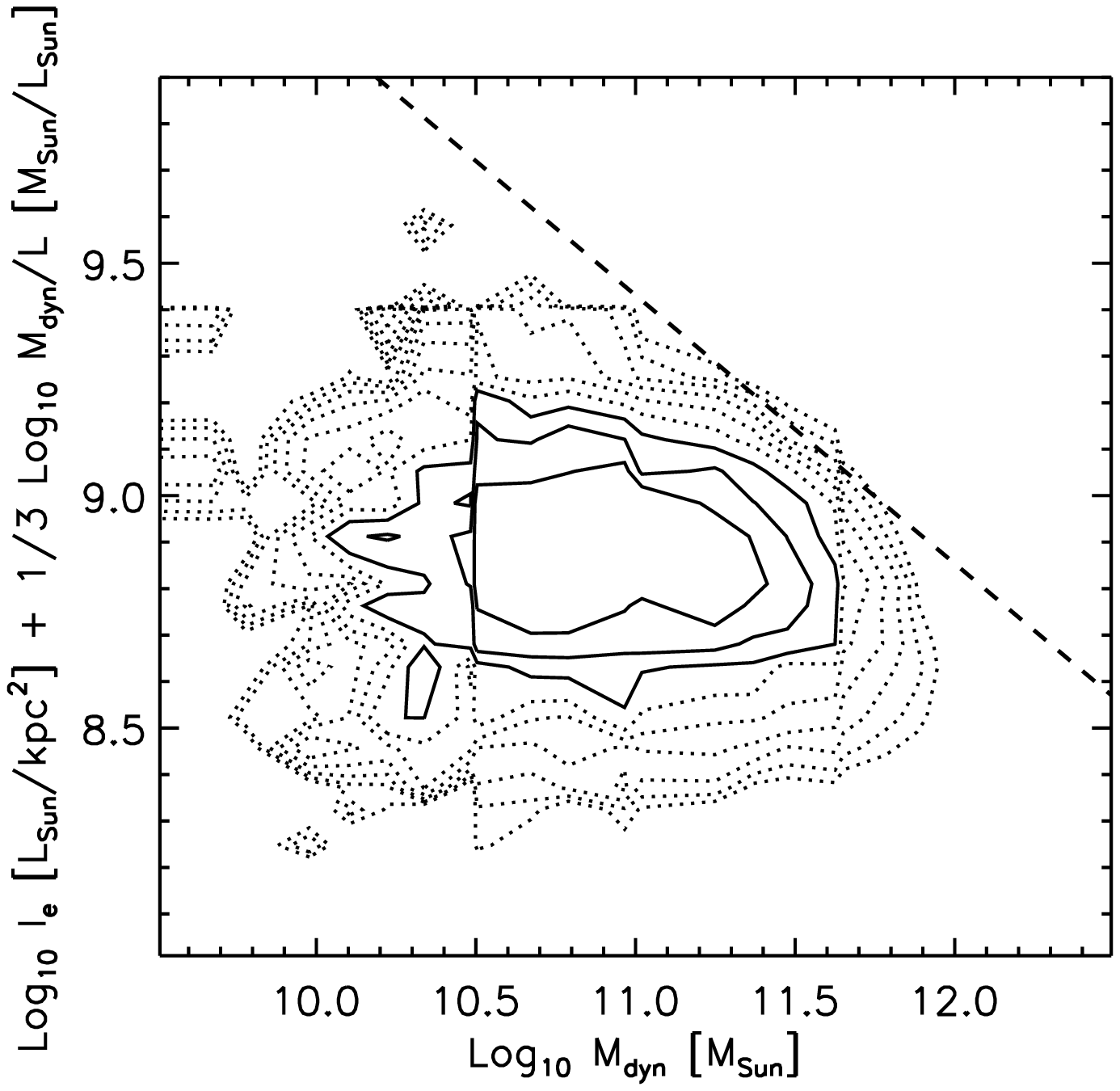}
 \includegraphics[width=0.475\hsize]{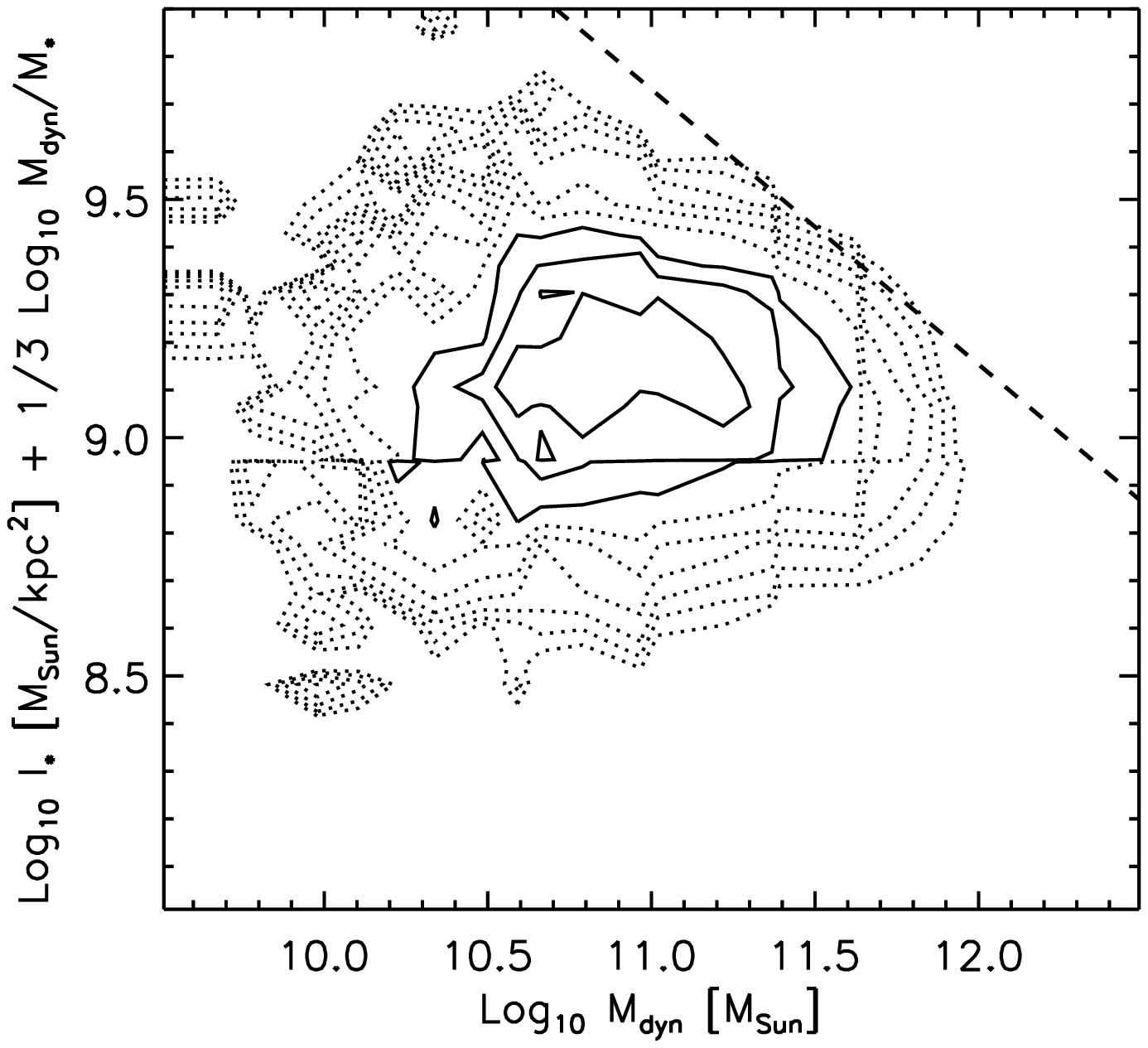}
 \caption{The $\kappa$-space Fundamental Plane in $r$ band (left panels) 
          and with stellar masses (from Gallazzi et al. 2005) 
          in place of luminosity (right panels).
          The figure shows $\kappa_1$, $\kappa_2$ and $\kappa_3$ transformed
          to solar units.
          Top panels show $\sqrt{3}\kappa_3$ versus $\sqrt{2}\kappa_1$; 
          dashed lines show the direct and inverse linear fits and 
          solid line the bisector one. 
          Bottom panels show $\sqrt{6}\kappa_2/3$ versus $\sqrt{2}\kappa_1$.
          Dashed line sloping down and to right shows 
          $\kappa_1+\kappa_2=$~constant scaled to solar units.}
 \label{FPkspace}
\end{figure*}

Figure~\ref{FPkspace} compares the two most interesting projections 
of $\kappa$- (left) and $\kappa_*$-space (right), when all variables 
have been expressed in solar units.  
The top panels show the `edge-on' view: $M_{dyn}/L-M_{dyn}$ and $M_{dyn}/M_*-M_{dyn}$.  
Note that the top right panel is essentially the same as 
Figure~\ref{showMs}, except that now we show the distribution 
using contours (objects have been weighted by $1/V_{\rm max}$ 
to account for selection effects); dashed lines show forward and 
inverse fits, and the solid line, which has slope 0.383 and 0.429 
in the two panels, shows the bisector fit.  

The bottom panels show the distribution of objects within the 
Plane.  The dashed line in the panel on the left shows 
 $\kappa_1 + \kappa_2 = $constant (as suggested by Bender et al. 1992).  
In the solar units used in the plot, the dashed line shows
\begin{equation}
 \left(\frac{I_e}{10^{10}L_\odot/{\rm kpc}^2}\right)
          \left(\frac{M_{dyn}/L}{M_\odot/L_\odot}\right)^{1/3}
           = 1.02\,\left(\frac{M_{dyn}}{10^{10}M_\odot}\right)^{-1/\sqrt{3}},
\end{equation}
whereas it is 
\begin{equation}
 \left(\frac{M_*/R_e^2}{10^{10}M_\odot/{\rm kpc}^2}\right)
          \left(\frac{M_{dyn}}{M_*}\right)^{1/3}
           = 2.04\,\left(\frac{M_{dyn}}{10^{10}M_\odot}\right)^{-1/\sqrt{3}}
\end{equation}
for the panel on the right.  
In the case of stellar masses, it is helpful to cube both sides 
of this expression, and then rearrange so that all the dependence 
on the dynamical mass is on the same side.  This yields 
\begin{equation}
 \left(\frac{M_*/R_e^3}{10^{10}M_\odot/{\rm kpc}^3}\right)^2
           = 8.49\,\left(\frac{M_{dyn}}{10^{10}M_\odot}\right)^{-\sqrt{3}-1};
\end{equation}
evidently, the dashed line expresses a relation between stellar 
density and dynamical mass:  the upper limit to the stellar density 
is approximately proportional to $M_{dyn}^{-4/3}$.

\section{Discussion and Conclusions}
We showed that the slope of the traditional, luminosity-based 
Fundamental Plane FP depends weakly but systematically on 
waveband -- it steepens towards the virial relation 
in the redder bands (Figure~\ref{FPgriz} and Table~\ref{griz}).  
Replacing $L$ with stellar mass $M_*$ leads to coefficients which 
are slightly closer to the virial ones; this is true whether one 
estimates $M_*/L$ from the spectra or simply from broad band colors 
(Table~\ref{abMst} and Figure~\ref{FPmstars}). 
 
The stellar mass FP is slightly thinner (with an average intrinsic scatter
of $\sim 0.048$~dex) than the luminosity-based FP (which has a typical
intrinsic scatter larger than $\sim 0.055$~dex).
The intrinsic scatter also decreases with wavelengths: from 0.062~dex in $g$ 
to 0.054~dex in $z$.  

We also showed that the intrinsic scatter around both the luminosity-based
and stellar mass Planes becomes significantly broader at 
low sizes/masses (Figures~\ref{residFPgriz} and 
\ref{FPmstars}) and that measurement errors account for essentially all 
of the observed scatter (about 0.04~dex) at large sizes/masses, 
suggesting that the Plane is rather thin for the very massive galaxies.

The fundamental nature of the stellar mass Plane FP$_*$, and the 
fact that $g-r$ color is a good indicator of $M_*/L$ 
(e.g. Bell et al. 2003), explains why residuals from FP correlate 
with color (e.g. Bernardi et al. 2003), or, equivalently, why color 
may be thought of as the fourth Fundamental parameter in early-type 
galaxy scaling relations.  

The fact that FP$_*$ does not quite have the virial scalings 
suggests that the ratio of stellar to dynamical mass, $M_*/M_{dyn}$ 
should vary across the population.  Figure~\ref{showMs} showed 
this was indeed the case:  $M_{dyn}/M_* \propto M_{dyn}^{0.17}$.  
This is in qualitative agreement with the results of the SAURON 
project (Cappellari et al. 2007), which is based on a very different 
analysis technique.  
At higher redshifts, the Fundamental Plane method is technically 
less challenging, so we expect it to provide a useful measure of 
the evolution of $M_{dyn}/M_*$.  This will also provide a useful 
check on the suggestion that the slope of the correlation between 
$M_{dyn}$ and $M_*$ does not evolve out to $z\sim 1$ (Bundy et al. 2007).  

We also presented an analysis of $\kappa$-space (Bender et al. 1992), 
but after replacing luminosities with stellar masses (using values from
Gallazzi et al. 2005).  
The plot of $M_*/M_{dyn}$ versus $M_{dyn}$ referred to above is 
also known as the edge-on view of $\kappa_*$-space.  
The face-on view of the Plane in this space 
(Figure~\ref{FPkspace}) showed that the maximum stellar density is 
smaller in the more massive objects:  $M_*/R_e^3 \propto M_{dyn}^{-4/3}$.  
The $\kappa_*$-space scalings at low redshift, and an estimate of 
how they evolve, should provide interesting constraints on galaxy 
formation models.  

Datasets are now sufficiently large that statistically significant 
curvature in most scaling relations has now been seen 
(e.g. Hyde \& Bernardi 2009).   
However, illustrating that the Fundamental Plane is warped is 
more difficult.  Previous work has shown that the coefficients 
of the FP can change dramatically as objects of different type 
(e.g. low luminosity or $\sigma$) are removed from the sample 
(e.g. Bernardi et al. 2003 for dependence on $\sigma$, and 
Donofrio et al. 2008 and Nigoche-Netro et al. 2008 for 
dependence on $L$).  
This has lead to the suggestion that these changes indicate that 
the FP is warped (e.g. Donofrio et al. 2008).  We showed that similar 
changes arise even in samples where the underlying pairwise scaling 
relations are not curved -- i.e., when the Plane is not warped
(Sections~\ref{cutl} and~\ref{cutv}).  Thus, the dependence of the 
fit parameters on the range of $L$ or $\sigma$ in the sample is not, 
by itself, evidence for curvature.  This dependence on the range 
of $L$ or $\sigma$ may explain some of the relatively wide range 
of Fundamental Plane coefficients in the literature.  

A more robust measure of how warped the Plane was also 
discussed.  We showed that a good understanding of the errors is 
necessary to interpret the results; else, correlated errors might 
lead one to conclude there is curvature even when there is none 
(Figure~\ref{residFPgriz} and discussion in Section~\ref{curved}).  
We conclude that the Plane is warped at the low size/mass end 
(Figures~\ref{residFPgriz} and~\ref{FPmstars}), and that it is 
also significantly broader at low sizes/masses.  These conclusions
are not affected by a low rate of spiral galaxy contamination, as
discussed in Appendix A.

Our analysis of the bias introduced in the FP coefficients by 
cuts in $L$ and $\sigma$ raises an interesting puzzle.  
The $\sim 50$ objects in the SLACS sample (Bolton et al. 2008) were 
selected because they are gravitational lenses, meaning that they 
tend to have large velocity dispersions.  The sample is particularly 
interesting because dynamic, stellar and gravitational lensing mass 
estimates are available.  Although the small sample size means the 
intrinsic thickness of the FP in this sample is almost certainly 
underestimated (Figure~\ref{err_rms}), what is of interest here 
is the fact that the smallest velocity dispersion in this sample 
is $\sigma \approx 160$~km~s$^{-1}$.  

Bolton et al. make no attempt to correct for the fact that 
their sample has no objects with small $\sigma$.  Instead, they  
argue that, given the same size and velocity dispersion 
distributions, the SLACS sample has approximately the same 
luminosity distribution as the parent SDSS sample from which 
it was drawn.  They also report that the coefficients of the 
Fundamental Plane for their sample are rather close to the virial 
scalings, and they use this to motivate a number of conclusions 
about the origin of `tilt' in most early-type samples.
However, our Figure~\ref{FPabV} suggests that if objects with 
$\sigma < 100$km~s$^{-1}$ are missing from the sample (this is 
a fairly standard cut-off based on what most spectrographs used 
for this work are capable of), and no account is made of this, 
then $a_{\rm orth}$ is biased high by about 15\% (also see 
discussion in Section 2.2 as well as Figure~4 in 
Bernardi et al. 2003).  
If objects with $\sigma<160$~km~s$^{-1}$ are missing, then 
$a_{\rm orth}$ is biased to values which are close to 2!  
Thus it is possible that the slope $a_{\rm orth}$ in the SLACS 
sample is high simply because it is biased.  If so, then the 
conclusions about the origin of `tilt' should be revised.

\section*{Acknowledgements}
We thank the organizers of the meeting held in Ensenada, Mexico 
in March 2008 for inviting us to attend, which prompted us to 
complete this work. We also thank R. Sheth for helpful discussions,
and A. Gallazzi for help in comparing different stellar mass estimates.
J.B.H. was supported in part by a Zaccaeus Daniels fellowship. 
J.B.H. and M.B. are grateful for additional support provided by 
NASA grant LTSA-NNG06GC19G. 

Funding for the Sloan Digital Sky Survey (SDSS) and SDSS-II Archive has been
provided by the Alfred P. Sloan Foundation, the Participating Institutions, the
National Science Foundation, the U.S. Department of Energy, the National
Aeronautics and Space Administration, the Japanese Monbukagakusho, and the Max
Planck Society, and the Higher Education Funding Council for England. The
SDSS Web site is http://www.sdss.org/.

The SDSS is managed by the Astrophysical Research Consortium (ARC) for the
Participating Institutions. The Participating Institutions are the American
Museum of Natural History, Astrophysical Institute Potsdam, University of Basel,
University of Cambridge, Case Western Reserve University, The University of
Chicago, Drexel University, Fermilab, the Institute for Advanced Study, the
Japan Participation Group, The Johns Hopkins University, the Joint Institute
for Nuclear Astrophysics, the Kavli Institute for Particle Astrophysics and
Cosmology, the Korean Scientist Group, the Chinese Academy of Sciences (LAMOST),
Los Alamos National Laboratory, the Max-Planck-Institute for Astronomy (MPIA),
the Max-Planck-Institute for Astrophysics (MPA), New Mexico State University,
Ohio State University, University of Pittsburgh, University of Portsmouth,
Princeton University, the United States Naval Observatory, and the University
of Washington.

\appendix

\section{Sample selection and disk contamination}
The main text describes how our sample was selected; the main cut 
is on the shape of the light profile ${\tt fracDev}=1$.  However, 
a non-negligible fraction of these objects (about 20\%) have 
$b/a<0.6$, which we remove because we believe the axis ratio 
is caused by an edge-on disk component.  Presumably, similar 
objects viewed faced-on remain in our sample, so the question 
arises as to how they may have affected our results.  

\begin{figure*}
 \centering
 \includegraphics[width=\hsize]{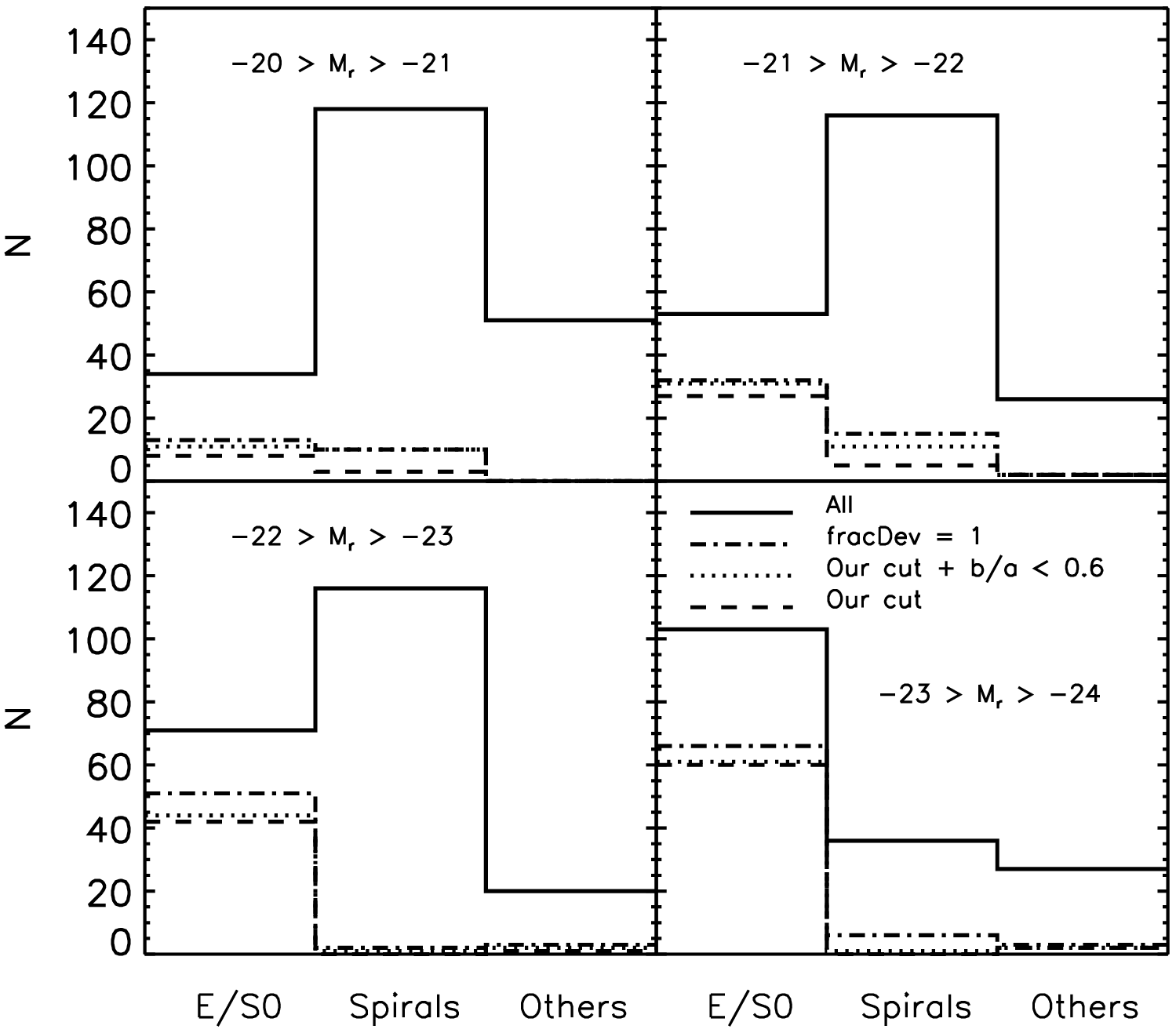}
 \caption{Effect of selection cuts on the morphological composition 
          in our sample, shown for four bins in absolute magnitude.  
          Solid histogram shows everything, dot-dashed is the 
          subset with {\tt fracDev}\ = 1, dotted is the smaller 
          subset which has spectra and satisfies our cuts in 
          {\tt eClass}, and dashed is for $b/a\ge 0.6$. }
 \label{morph}
\end{figure*}

To address this, we divided the full sample in 5 bins in 
luminosity, randomly selected $\sim 200$ galaxies in each bin, and 
classified them, by eye, as E/S0, Spirals, and Others (irregulars, 
low surface brightness objects, interacting systems, projections).  
The different panels in Figure~\ref{morph} show these classifications in four  
luminosity bins.  In each panel, the solid histogram (the one 
with the most counts) contains $\sim 200$ objects (a few objects were
removed because of contamination of a bright star or misclassification).  
The dot-dashed histogram shows the subset with ${\tt fracDev = 1}$ in 
both $g$ and $r$ bands, the dashed histogram shows the result of applying all 
our other cuts except the one on $b/a$, and the dotted shows the result of 
excluding objects with $b/a < 0.6$.  This shows that 
${\tt fracDev = 1}$ removes most of the Spirals and Others -- 
our further cuts help in reducing the left over contamination.  

\begin{figure*}
 \centering
 \includegraphics[width=\hsize]{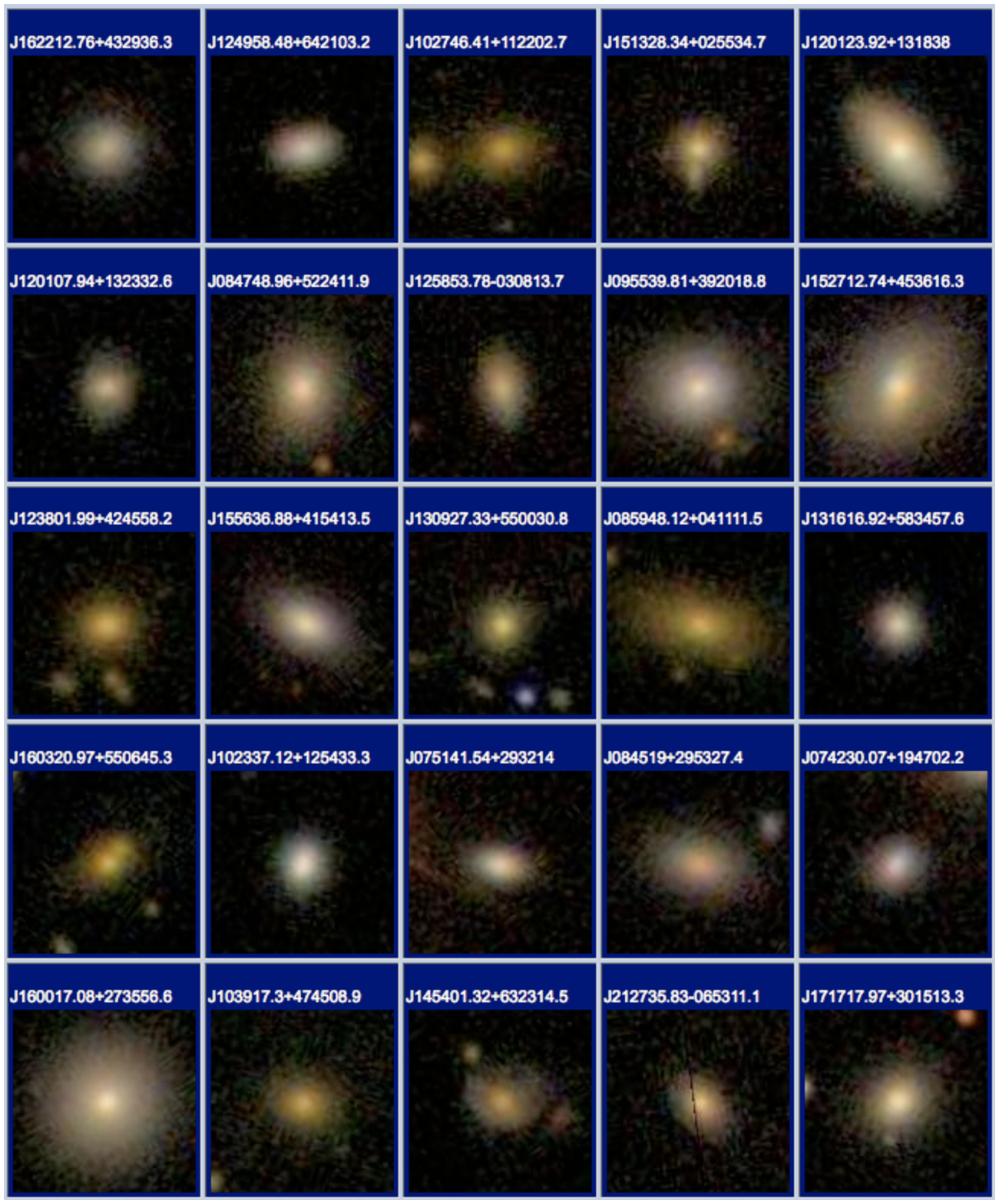}
 \caption{Selection of objects classified by eye as E/S0s, but which 
          have ${\tt fracDev < 1}$.  Objects such as these are shown 
          by the diamond symbols in Figure~\ref{contaminatedFP}. }
 \label{notEs}
\end{figure*}

\begin{figure*}
 \centering
 \includegraphics[width=0.7\hsize]{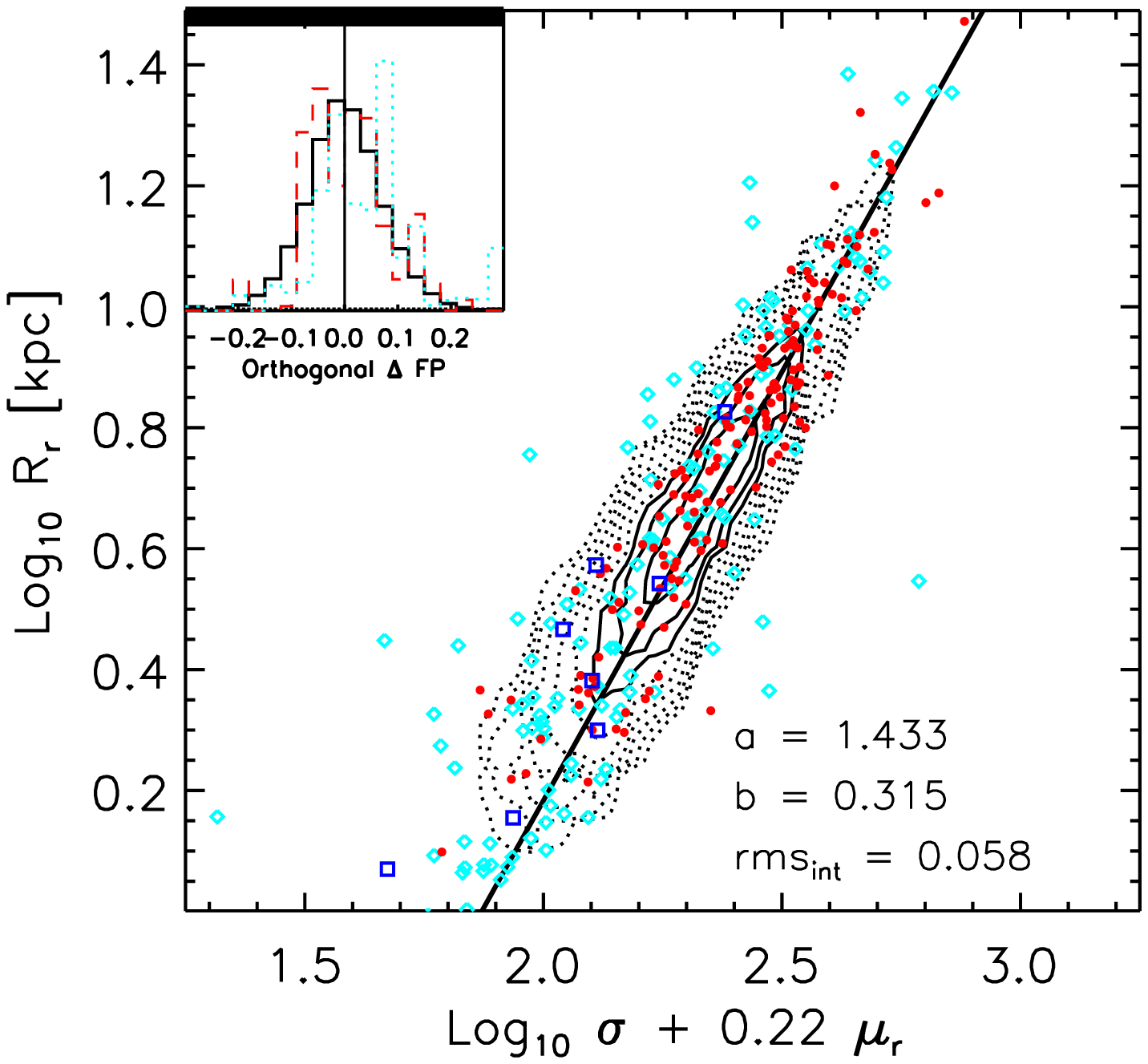}
 \caption{The Fundamental Plane in the $r$ band 
          (same as top right panel in Figure~\ref{FPgriz}).
          Filled circles and open squares show objects which satisfied our 
          selection cuts and were morphologically classified by eye as
          E/S0 (filled circles, see also dashed line in the inset at top left) 
          and Spirals (open squares). 
          Diamond-shaped symbols (and dotted line in the inset at top left)
          show objects we classified as E/S0 by eye 
          but do not have ${\tt fracDev = 1}$.}
 \label{contaminatedFP}
\end{figure*}

While our selection removes most of the later type galaxies, it is 
perhaps surprising that a significant fraction of the objects 
we classified as ellipticals by eye do not have ${\tt fracDev = 1}$.  
A selection of these objects is shown in Figure~\ref{notEs}; it is 
evident that many of these objects are only marginally E/S0s, 
so we are confident that ${\tt fracDev = 1}$ is a reliable cut.  
To illustrate the effect that such objects might have had on our 
results, the diamond-shaped symbols in Figure~\ref{contaminatedFP} 
show their location relative to the FP defined in the main text.  
These objects are offset to higher sizes, and they have 
substantially higher scatter around the FP (see inset at top left), 
suggesting that our decision to exclude them from the analysis in 
the main text, is reasonable.  

The filled circles and open squares in Figure~\ref{contaminatedFP} 
show objects we classified as E/S0s and Spirals, and which satisfied 
our selection cuts on ${\tt fracDev}$ and $b/a$ (i.e., they make up 
the dotted histogram in Figure~\ref{morph}).  There are a handfull of 
spirals and they tend to have larger $R_e$ -- as one might expect from 
trying to fit a disk component with a single deVaucouleur profile.  
They are a slightly larger fraction of the total counts at small 
$\sigma + 0.21\mu_e$, however they are so few that they do not
bias the slope, scatter, or curvature properties of the FP. Additionally, the curvature
described in Section~\ref{curved} results in negative residuals at
low $X_{FP}$. Any curvature introduced by spiral contamination would introduce
positive residuals because of artificially large radii.

\label{lastpage}

\end{document}